\documentclass[fleqn,usenatbib]{mnras}

\usepackage{newtxtext,newtxmath}
\usepackage[T1]{fontenc}

\DeclareRobustCommand{\VAN}[3]{#2}
\let\VANthebibliography\thebibliography
\def\thebibliography{\DeclareRobustCommand{\VAN}[3]{##3}\VANthebibliography}

\usepackage{graphicx}	% Including figure files
\usepackage{amsmath}	% Advanced maths commands

\usepackage{makecell}
\newcommand{\tess}{\textit{TESS}}
\newcommand{\exofastv}{{\tt EXOFASTv2~}}
\newcommand{\multifast}{{\tt MultiFast~}}
\newcommand{\exofast}{{\tt EXOFAST~}}
\newcommand{\teff}{$T_{\rm eff}$}

\usepackage{tabularx}
\usepackage{graphicx,caption}
\usepackage{soul}
\usepackage{booktabs}
\usepackage{pdflscape}
\newcommand{\bjdtdb}{\ensuremath{\rm {BJD_{TDB}}}}

\newcommand{\msun}{\ensuremath{\,M_\odot}}
\newcommand{\rsun}{\ensuremath{\,R_\odot}}
\newcommand{\lsun}{\ensuremath{\,L_\odot}}

\newcommand{\mj}{\ensuremath{\,M_{\rm J}}}
\newcommand{\rj}{\ensuremath{\,R_{\rm J}}}

\newcommand{\fave}{\langle F \rangle}
\newcommand{\fluxcgs}{10$^9$ erg s$^{-1}$ cm$^{-2}$}

\newcommand{\specialcell}[2][c]{%
  \begin{tabular}[#1]{@{}l@{}}#2\end{tabular}}
\title[Examining Systematics in KELT-15]{Exploring Systematic Errors in the Inferred Parameters of the Transiting Planet KELT-15b and its Host Star}

\author[A. Duck et al.]{
Alison Duck,$^{1}$$^{\href{https://orcid.org/0000-0002-4531-6899}{\includegraphics[scale=0.5]{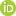}}}$,
B. Scott Gaudi,$^{1}$$^{\href{https://orcid.org/0000-0003-0395-9869}{\includegraphics[scale=0.5]{orcid.jpg}}}$,
Jason D. Eastman$^{2}$$^{\href{https://orcid.org/0000-0003-3773-5142}{\includegraphics[scale=0.5]{orcid.jpg}}}$,
Joseph E. Rodriguez$^{3}$$^{\href{https://orcid.org/0000-0001-8812-0565}{\includegraphics[scale=0.5]{orcid.jpg}}}$
\\
$^{*}$duck.18@osu.edu\\
$^{1}$Department of Astronomy, The Ohio State University, Columbus, OH 43210, USA\\
$^{2}$Center for Astrophysics \textbar \ Harvard \& Smithsonian, 60 Garden St, Cambridge, MA 02138, USA\\
$^{3}$Center for Data Intensive and Time Domain Astronomy, Department of Physics and Astronomy, Michigan State University, East Lansing, MI 48824, USA
}

\date{Accepted XXX. Received YYY; in original form ZZZ}

\pubyear{\the\year{}}

\begin{document}
\label{firstpage}
\pagerange{\pageref{firstpage}--\pageref{lastpage}}
\maketitle

%%%%%%%%%%%%%%%%%%%%%%%%%%%%%%%%%%%%%%%%%%%%%%%%%%%%%%%%%%%%%%%%%%
\begin{abstract}

Transiting planet systems offer a unique opportunity to measure the masses and radii of many planets and their host stars. Yet, relative photometry and radial velocity measurements alone only constrain the density of the host star. In remedy, the community uses theoretical and semi-empirical methods to break this one-parameter degeneracy and measure the mass and radius of the host star and its planet(s). We investigate the differences in the inferred system parameters due to modeling a host star with the Torres mass-radius relations, YY evolutionary tracks, MIST evolutionary tracks, and a stellar radius estimate from the spectral energy distribution (SED). We consider the effects of different priors on the stellar effective temperature, limb darkening, and eccentricity of the planet. Using the publicly available software package {\tt EXOFASTv2}, we globally model TESS photometry and radial velocity observations of KELT-15, which hosts a fairly representative hot Jupiter. In total, we explore the impact of 28 different choices of priors on the inferred parameters of KELT-15b. We find broad agreement in the inferred system parameters across methodologies at the level of $\sim 1.1 \sigma$ between the MIST and SED constraints. This gives some confidence that systematic errors are not ubiquitous in transiting planets systems. We also find a $\sim 2 \sigma$ difference in the $R_*$ estimated by the MIST models when we adopt differing literature spectroscopic \teff\ estimates. Similar studies of a large number of systems are needed to definitely assess systematic uncertainties the exoplanet population as a whole.

\end{abstract}


%%%%%%%%%%%%%%%%%%%%%%%%%%%%%%%%%%%%%%%%%%%%%%%%%%%%%%%%%%%%%%%%%%
\section{Introduction} \label{sec:intro}

Accurate masses and radii are generally needed to determine the basic nature of exoplanets. For example, estimates of the densities of exoplanets from their measured masses and radii have led to the discovery of previously unknown classes of planets such as ``Super Earths," ``Sub-Neptunes," and ultra low-density ``Puffballs." Indeed, the density determination of the first transiting planet detected, HD209458b \citep{Charbonneau2000} showed conclusively that Hot Jupiters such as 51 Peg b \citep{MayorQueloz1995} were planets with densities similar to the giant planets in our solar system. Moreover, rigorous comparative exoplanetology requires precise and accurate planetary masses and radii with a detailed understanding of the inherent systematic uncertainties to improve our understanding of the formation and evolution of planetary systems. Our goal is to take a first step in quantifying the systematic uncertainties in the inferred parameters of transiting planets arising from the choice of priors adopted when modeling these systems.

Here, we primarily focus on one particular choice of prior that could give rise to systematic errors, namely the method by which the mass-radius degeneracy in transiting planet system is broken. Even with arbitrarily precise measurements of the relative photometry and radial velocity (RV) of a transiting planet system, it is not possible to independently determine the mass $M_\star$ and radius $R_\star$ of the host star or the mass $M_{P}$ and radius $R_{P}$ of its transiting planet(s), as these measurements alone do not constrain the absolute mass or size scale of the system. Rather, these measurements can only constrain the density of the star, $\rho_{*}$ \citep{Seager_2003}, resulting in a one-parameter degeneracy. We must introduce additional information to set the scale of the system and measure the absolute masses and radii of the host star and planets. However, such additional information can introduce unquantified systematic uncertainties. We assess the magnitude of these systematic uncertainties my comparing the system parameters inferred using four different commonly-used methods of breaking the mass-radius degeneracy. A detailed discussion of this degeneracy can be found in Appendix \ref{app:A}.

We focus on one system, KELT-15b \citep{kelt15b}, a fairly typical hot Jupiter in terms of both its physical properties and the amount of observations available for the system. We perform a large suite of fits to these data, variously adopting or dropping different priors that are commonly considered when fitting transiting planets in 28 total iterations. For each fit, we find both the directly observable and inferred physical parameters of the system, and compare the differences in these parameters found for different choices of priors to the nominal statistical uncertainties.

In this work, we will analyze KELT-15b with \exofastv \citep{exofastv2,exofast1}. \exofastv  simultaneously fits photometric and radial velocity observations, and implements a variety of additional constraints to characterize the host star. We will explore the systematic uncertainties in the inferred system parameters due to (1) the choice of the method to break mass-radius degeneracy, (2) allowing for free eccentricity versus fixing the eccentricity to zero, (3) adopting an uninformative prior on the stellar effective temperature versus adopting a prior based on high-resolution spectra, and (4) adopting priors for the limb-darkening coefficients based on the properties of the host star versus fitting for these explicitly with an uninformative prior.

%Specifically, we consider four methods of characterizing the host star in order to break the mass-radius degeneracy. \exofastv allows the user to adopt external constraints from the Spectral Energy Distribution (SED) of the star, which when combined with the parallax and a constraint on the effective temperature, \teff\ (either from the SED itself or from, e.g., high-resolution spectroscopy), yields the stellar radius \citep{exofast1,exofastv2}. \exofastv also supports the Torres empirical mass-radius relations \citep{Torres_2009}, Yonsei-Yale evolutionary tracks \citep{YY2001}, and the Mesa Isochrone and Stellar Track (MIST) models \citep{mist} as methods of breaking the host star mass-radius degeneracy. These constitute the four methods examined in this study, which we will compare in the context of a transiting exoplanet system for the first time.

We consider four methods of characterizing the host star implemented in \exofastv to break the mass-radius degeneracy.  The first uses a measurement of $R_*$ from the spectral energy distribution (SED) of the star combined with a parallax and a constraint on the stellar effective temperature (either from the SED itself or from, e.g., high-resolution spectroscopy).  The second method adopts the empirical relations between $M_*$, $R_*$, $T_{\rm eff}$ and the surface gravity $\log{g}$ derived from main-sequence binaries by \citet{Torres_2009}.  The last two methods adopt the Yonsei-Yale (YY) evolutionary models by \citet{YY2001} and Mesa Isochrone and Stellar Track (MIST) evolutionary models by \citet{mist}. A qualitative description of how these four methods break the $M_*-R_*$ degeneracy can be found in Section \ref{sec:StellarModels} and Figure \ref{fig:StellarConstraints}.

The plan for this paper is as follows. In \S \ref{sec:Exofast}, we discuss \exofastv and its approach to globally modeling planetary systems and their host stars. In \S \ref{sec:StellarModels}, we summarize the four methods of breaking the host star mass-radius degeneracy which we will use to characterise KELT-15b and explore the information content provided by these methods. In \S \ref{sec:kelt15b}, we provide the rationale for selecting KELT-15b as the test-bed of this analysis. In \S \ref{sec:analysis}, we outline our analysis approach to systematically model KELT-15b. In \S \ref{sec:SanityCheck}, we check the robustness of our procedure by reproducing the results from the original analysis of this system from \cite{kelt15b}.  In \S \ref{sec:results}, we describe the results of our analysis. In \S \ref{sec:discussion}, we discuss the implications of these results on modeling transiting planets. Finally, in \S \ref{sec:conclusions}, we summarize our findings and their impact.

%%%%%%%%%%%%%%%%%%%%%%%%%%%%%%%%%%%%%%%%%%%%%%%%%%%%%%%%%%%%%%%%%%
\section{Description of \exofastv} \label{sec:Exofast}

We use the publicly-available code \exofastv \citep{exofastv2,exofast1} to model the KELT-15 system. EXOFASTv2 simultaneously and self-consistently fits the relative photometry and radial velocity data of a transiting planet, and, if desired, the spectral energy distribution of the host star. Notably, \exofastv allows one to adopt priors on the properties of the host star based on empirical fitting formulae (e.g., Torres) or stellar evolutionary models (YY, MIST).

The minimization and parameter estimation techniques are based on a Markov Chain Monte Carlo (MCMC) algorithm. At each step, \exofastv calculates the likelihood that a test model could produce the given observations. \exofastv uses the Gelman-Rubin statistic \citep{GelmanRubin} to estimate whether the chains have converged. The Gelman-Rubin statistic compares the variance of the mean of any given parameter across the individual chains to the mean of the variances. When the Gelman-Rubin statistic is close to unity, the chains are generally considered well-mixed, whereas values that are much greater than one are an indication that the chains have not converged. The Gelman-Rubin statistic also shows that after a certain point, adding steps to the chains no longer makes the chains appreciably more similar. Thus, the fitting procedure can be considered converged. By default, \exofastv uses a Gelman-Rubin statistic cut off of 1.01 \citep{exofastv2}. This cut off is stricter than the originally proposed cut off of 1.2 in \citet{GelmanRubin}.

\exofastv affords the user a wide array of customizable features. Given the user-supplied data for the system such as the light curve, radial velocity, Spectral Energy Distribution (SED), and parallax, \exofastv provides a menu of methods for carrying out the fit to the system, an in particular characterizing the host star.  Priors are naturally accounted for within {\tt EXOFASTv2}. For example, for a Gaussian prior the user supplies a mean value and a $1\sigma$ Gaussian uncertainty. Alternatively, the user can supply a uniform prior with a mean value and an upper and lower bound. If the user does not define an uncertainty in conjunction with their mean value for a given parameter, \exofastv applies a uniform unbounded prior on the parameter. Here we use {\tt EXOFASTv2}'s functionality to consider the impact of a prior on limb darkening coefficients based on the Claret limb darkening tables \citep{claret2011,claret2017} versus using an uninformative uniform prior. We also consider the impact of assuming a circular orbit versus allowing for free eccentricity. See \citet{exofastv2} for a description of the full functionality of {\tt EXOFASTv2}. 

{\tt EXOFASTv2}'s ability to fit all the data simultaneously and self-consistently allows for robust and precise estimates of the physical parameters of a transiting system.  In the past, researchers have sometimes adopted a piece-wise approach, whereby the properties of the host stars are first estimated based on, e.g., their effective temperatures, parallax, broad-band photometry, without reference to the light curve data. Then, the properties of the planet are determined by fitting light curve and radial velocity observations, adopting priors on the properties of the star inferred from the first step. However, this methodology does not account for the fact that the light curve and radial velocity data independently constrain the properties of the host star via $\rm \rho_{*}$ as shown in Eq. \ref{eqn:rhostar}. Information from light curve and radial velocity observations are consistently included, by globally modelling systems with \exofastv \citep{exofastv2}.  

Recently, \citet{eastman2023} demonstrated the power of this approach. They find that, even after accounting for the systematic uncertainties in the stellar models described in \citet{tayar2022}, they can use the stellar density constraint to determine stellar radii to 3\% and temperature to 1.75\% in a typical case.

%%%%%%%%%%%%%%%%%%%%%%%%%%%%%%%%%%%%%%%%%%%%%%%%%%%%%%%%%%%%%%%%%%
\section{Common Stellar Models and Assumptions}\label{sec:StellarModels}

Several sets of constraints are commonly used to break the mass-radius degeneracy in the host star in published discoveries of transiting planets. Thus the current ensemble of known transiting planets represent a sample whose properties have been inferred in an inhomogenous way. To date, few studies exist comparing the parameters inferred using different methods of breaking the degeneracy \citep{tayar2022}.  Therefore, while we may know the statistical precision of the reported mass and radius measurements of transiting planets, we generally do not know the accuracy and systematic uncertainty of these parameters. We particularly do not know whether the systematic uncertainties exceed the quoted precision in some subset of cases. 

\begin{figure*}
    \centering
    \includegraphics[width=8.5cm]{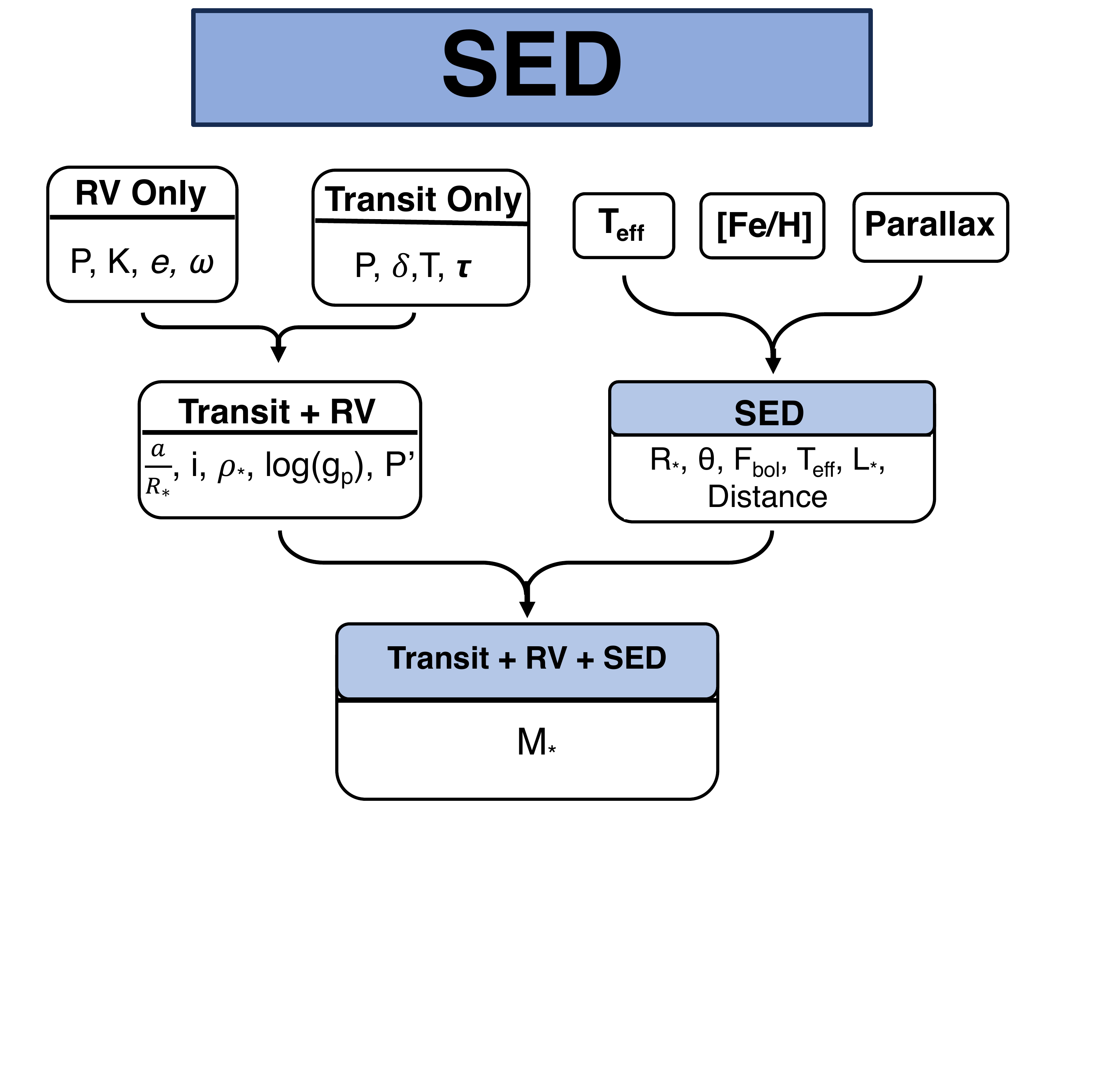}
    \includegraphics[width=8.5cm]{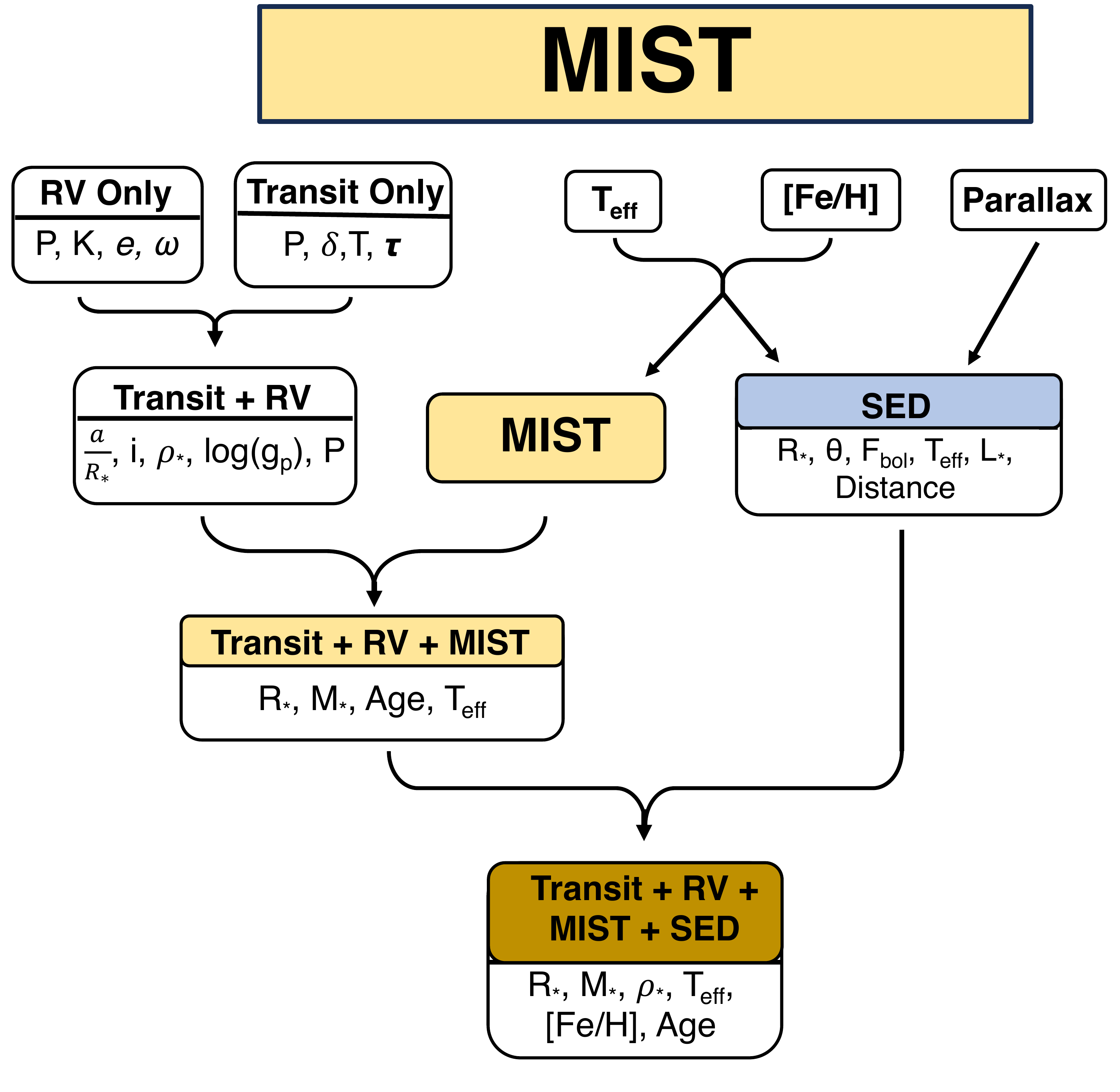}

    \caption{We graphically represent the information provided by each set of constraints on our host star and planetary system. As the constraints increase each step inherits information from the previous steps. We show selected parameters of interest that are constrained in each step. The YY and Torres stellar constrains provide similar sets of information as those provided by MIST. However, the Torres relations provide no information about the stellar effective temperature or stellar age. Though not represented the SED constraint does provide weak information regarding [Fe/H] and $\rm \log(g_{*})$. The boxes showing Transit + RV + SED and Transit + RV + MIST both represent our single constraint iterations. The Transit + RV + MIST + SED box represents a double constraint iteration which crucially provides detailed estimates of $R_{*}$, $M_{*}$, $\rho_{*}$, and \teff. }\label{fig:StellarConstraints}
\end{figure*}

We explore four frequently used constraints to model the host stars implemented in {\tt EXOFASTv2}. Figure \ref{fig:StellarConstraints} provides a schematic overview of the information provided by each set of data and model constraints on the properties of the host star and planetary system. One approach is applying semi-empirical relationships, such as the Torres relations \citep{Torres_2009}. This study looked at 95 detached double-lined eclipsing binary systems (as well as the $\alpha$ Centauri system) and found empirical polynomial relationships between the mass and radii of these stars and $\log{g_*}$, \teff, and [Fe/H]. As is well known, double-lined eclipsing binary systems allow for accurate measurements of the mass and radii of both stars. The Torres relations were determined by fitting primarily to unevolved or somewhat evolved main sequence stars. Thus they do not apply to giants or pre main sequence stars. Furthermore, only a handful of the stars in the \cite{Torres_2009} sample were low-mass M stars, and thus one should be wary of applying them to low-mass stars. Note that the Torres relations yielded masses and radii that are accurate to 6\% and 3\%, respectively, based on the scatter of the measured values of these quantities relative to those predicted by the relations. With measurements of $\log{g_*}$, \teff, and [Fe/H], these relations can be used to quickly estimate $M_*$ and $R_*$\footnote{\citet{Enoch_2010} fit the Torres data to find polynomial relationships between $M_*$ and $R_*$ and $\rho_*$, \teff, and [Fe/H], since $\rho_*$ is a direct observable in single-lined transiting planet systems. This is effectively done ``internally" by {\tt EXOFASTv2}, and thus we simply refer to the Torres relations.}.
  
Stellar Evolutionary tracks are a popular indirect method of estimating stellar properties. The Yonsei-Yale (YY) stellar evolutionary tracks followed the evolution of stars starting from the pre-main sequence. The YY tracks predict the luminosity, color, \teff, and radius as a function of mass, age, and metallicity of the star \citep{YY2001}. We note that the YY evolutionary tracks used in \exofastv do not apply to stars less than $0.5~M_\odot$ \citep{exofastv2}. In this study, the mass of our host star is roughly $1.2~M_\odot$, thus in the range where the YY evolutionary tracks are applicable.

The MESA Isochrone and Stellar Tracks (MIST) models \citep{mist} are the default stellar evolutionary model for \exofastv \citep{exofastv2}. The MIST models build off the Modules for Experiments in Stellar Astrophysics (MESA) framework \citep{Paxton2011,Paxton2013,Paxton2015} resulting in a set of comprehensive stellar evolutionary tracks and isochrones. The MIST evolutionary tracks are available for stars between $0.1~M_\odot$ to $300~M_\odot$, starting at 100,000 years in age, thus including pre-main sequence stars \citep{mist}. This range includes nearly any planet hosting star, making the framework the most flexible of the theoretical models examined here. MIST computes the stellar evolutionary tracks using a grid of initial mass, initial [Fe/H], and evolutionary phase \citep{exofastv2}. 

For the single constraint fits using the (MIST, YY, Torres) models, $\rho_*$ is constrained solely from the light curve and radial velocity data through Eqs. \ref{eqn:aoverrstar} and \ref{eqn:rhostar} in Appendix \ref{app:A}.  When combined with a prior on  \teff\ and [Fe/H], the models then constrain $R_*$ and $M_*$.  For the Torres models, this is done through the Torres relations \citep{Torres_2009}, which parameterize $R_*$ and $M_*$ as polynomial functions of \teff, [Fe/H], and $\log{g_*}$. These relations can be rewritten to be a function of \teff, [Fe/H], and $\rho_*$ \citep{Enoch_2010}. In essence, \teff\ and [Fe/H] parameterize the location of the star on the zero age main sequence (ZAMS), and $\rho_*$ parameterizes how far the star has evolved off of ZAMS.  For the MIST and YY evolutionary tracks the situation is similar, except the relations are not simple polynomials, and age becomes a proxy for the evolutionary state of the star and thus $\rho_*$.  Therefore, the single constraint fits with the Torres relations do not yield ages, whereas those with the evolutionary tracks do.

The most empirical of these techniques is the method of fitting a model Spectral Energy Distribution (SED) to absolute broad band photometry measurements of the host star. The SED fitting technique is mostly independent of stellar models \citep{stassun2016}. In practice, SED fitting integrates over stellar atmospheric models making it not fully empirical and subject to the biases of such models. Still, the SED can be used in conjunction with MIST, YY, or Torres constraints without double counting information about the star.  Here, $\rho_*$ is again constrained solely by the light curve (which constrains the Period, $\delta$, $T$, and $\tau$) and radial velocity data (which constrains the Period, $\rm K_{*}$, $e$, $\omega$). SED fitting places constraints on \teff, bolometric luminosity $F_{\rm bol}$, and the extinction $A_V$. EXOFASTv2 suggests that users constrain $A_V$ using upper limits from \citep{Schlafly}. Then $R_*$ is constrained via $F_{\rm bol}$ and \teff, and the parallax from Gaia. We correct the \textit{Gaia} DR3 parallax with the systematic offset described by \cite{Lindegren2021}. In practice, the SED reports larger uncertainties in \teff\ than using spectroscopic priors, but the constraint on $\rho_{*}$ from the transit \citep{eastman2023} and a proper accounting for $2\%$ error floors in spectroscopic \teff\ measurements \citet{tayar2022}, makes SED competitive with other methods. Still, the uncertainty in $R_*$ is reduced by imposing a spectroscopic \teff\ constraint when combining the estimate of $F_{\rm bol}$ with the Gaia parallax. A spectroscopic prior constraining [Fe/H] provides a similar improvement in the precision of the $F_{\rm bol}$ estimate. In principle, the model spectrum fit to the SED and used to infer $F_{\rm bol}$ also depends on $\log{g_*}$, but this dependence is weak and so doesn't qualitatively effect the previous argument. SED fitting should not be used if there is suspicion that its broad band photometry is blended with another star.

In all four of these instances, the constraint on ${\rho_*}$ from the light curve is needed to make independent measurements of $M_*$ and $R_*$, and thus a complete solution to the system. Since the priors on \teff\ and [Fe/H] are the same in all the `default' single constraint models, and the only constraint on $\rho_*$ comes from the light curve, we expect agreement in stellar density across the single constraint models, as observed. 

Increasingly, combining theoretical models with an SED derived from multi-wavelength observations, is becoming a common approach. As described, the SED can provide additional constraints on the radius or effective temperature of the host star. When combined with a theoretical model like MIST, this can further narrow the parameter space of plausible values of host star mass and radius. Combining these constraints necessitates an understanding of the systematic uncertainties in both components. This is vital to understanding the complete error budget of derived stellar and planetary properties.

For such double constraint fits, the system is over-constrained.  Conceptually, it is tempting to think of this in the following manner: the (MIST, YY, Torres) models combined with the constraint on $\rho_*$ from light curve and the radial velocity data, and the \teff\ and [Fe/H] priors, yield $R_*$ and $M_*$ and, in the case of the MIST and YY fits, also the age.  The SED+parallax+\teff\ prior+[Fe/H] prior yield $R_*$.  Thus, there are two independent ways in which $R_*$ is constrained, and the the double constraint fits should therefore yield a value of $R_*$ that is essentially an average of the two values weighted by their uncertainty, and the value of $M_*$ would be unchanged.  Indeed, this might be the result if the light curve and RV data were fit independently of SED+parallax data, as has been occasionally been done in the literature.

In fact, all of the data must be analyzed simultaneously. This has several effects. First, the values of both $R_*$ and $M_*$ inferred from the joint fit will differ from those inferred from the single constraint fits. Since the fit is simultaneously applying both constraints, the fits will inform each other as opposed to averaging two wholly independent estimates. Second, other parameters inferred from the fits, such as the age, will also be different in the double constraint fits as compared to the single constraint fits. Now the age estimate will also be incorporating information on the \teff\ and luminosity from the SED fit while applying the constraints from the stellar evolution model. Finally, since both $R_*$ and $M_*$ differ, so does $\rho_*$. This immediately implies that the values of the `direct observables' such as $T$, $\tau$, and $\delta$ can differ as well.

\subsection{Other methods of stellar characterization}\label{sec:Variation}
These four methods are not the only ones currently in use. Spectroscopic stellar surface gravity measurements can provide a direct estimates of $\log{g_*}$ and thus be used in conjunction with the transit and RV data to derive $M_{*}$ and $R_{*}$. However, these observations require high signal-to-noise ratio, high resolution (R$\ge$ 50,000) spectra, and thus can be prohibitively resource intensive. The estimates of $\log{g_*}$ produced by this method are often neither very precise nor very accurate \citep{Heiter2015,Mortier2014,2010A&ARv..18...67T}. These values can also be estimated  assuming we have spectroscopic \teff\ or [Fe/H], or radial velocity observations. Flicker-based surface gravity measurements can place another constraint on $\log{g_*}$. Granulation on the surface of stars can produce variations in flux, or flicker, which is strongly correlated with $\log{g_*}$ \citep{Bastien2013,Bastien2016}. However, the relationship between $\log{g_*}$ and $R_{*}$ requires calibration and thus these measurements do not strictly apply to individual systems.

Asteroseismology can provide another avenue of estimating $R_{*}$ and $M_{*}$. Acoustic oscillations in stars produce a power spectrum of brightness fluctuations. The large frequency spacing of the peaks in the power spectrum $\Delta v$ is directly related to $\rho_{*}$, whereas the frequency of the maximum power $\nu_{\rm max}$, is related to $\log{g_*}$ and \teff\ via an empirically calibrated scaling relation \citep{Kjeldsen1995}. These parameters can be combined for a semi-empirical estimate of $\rm R_{*}$ and $\rm M_{*}$. 

However, both asteroseismology and flicker measurements are observationally expensive methods. They require high-cadence, high-precision light curve observations with long baselines such as those from Kepler \citep{Borucki2010}. The Kepler mission had great success in applying asteroseismic techniques to is planet hosting stars \citep{huber2013, lundkvist2016, vaneylen2018,chontos2019}. However, despite \tess\ discovering thousands of planet candidates, asteroseismology has been applied to relatively few host stars \citep{huber2022, Jiang2020}. Therefore, it would be difficult to obtain asteroseismic observations for wide swaths of the current population of planet hosting stars. Thus these methods were considered to be beyond the scope of this study.

\subsection{Priors on the eccentricity, limb-darkening, and \teff} \label{sec:StellarAssumptions}

In addition to the choice of methods to break the mass-radius degeneracy, we consider the effect of adopting different priors on the effective temperature of the star, the eccentricity of the orbit, and the limb-darkening of the star. The specific choice for these priors can also introduce systematic uncertainties in the inferred parameters of the system, which we explore here.

It is commonly assumed, in the absence of a definitive detection of non-zero eccentricity, that the orbits of hot Jupiters are exactly circular. However, this assumption may not strictly hold for all such systems, and therefore we consider the impact of relaxing this prior on the inferred parameters and their uncertainties.  In Equation \ref{eqn:K} in Appendix \ref{app:A}, we see that $K_{*}$, the semi-amplitude in radial velocity observations, depends on $e$. Thus the derived ratio $M_{p}/M_{*}$ consequently shares a dependence on eccentricity. By fixing $e$ at zero, we remove any uncertainty in eccentricity that would contribute to the total error budget in $M_{p}$, to the extent that $e$ is correlated with $K_*$. This could be artificially constraining the error in planet mass by assuming $e=0$ ab initio. 

Orbital eccentricity assumptions also impact our knowledge of stellar properties. From Equations \ref{eqn:aoverrstar} and \ref{eqn:rhostar} in Appendix \ref{app:A}, we see that the inferred density of the star depends on the eccentricity and argument of periastron. By examining $\frac{a}{R_*}$ in Eq. \ref{eqn:aoverrstar}, we see that the ratio depends on eccentricity. When $\frac{a}{R_*}$ is used in Eq. \ref{eqn:rhostar} to constrain the density, that dependence on eccentricity follows. By assuming $e = 0$, it is possible that this is over-constraining the uncertainty in the derived stellar density. Thus, eccentricity assumptions should be well tested due to their impacts on derived parameters regarding both the host star and the planet.

%%%%%%%%%%%%%%%%%%

Explicit assumptions are also required when considering estimates of stellar limb darkening coefficients. It is also possible within the framework of \exofastv to adopt priors on the stellar limb darkening coefficients using the Claret limb darkening tables \citep{claret2011,claret2017}. The Claret limb darkening coefficients are tabulated in a grid of filters, \teff, $\log(g)$, and [Fe/H] from the PHOENIX \citep{pheonix2009,pheonix} and ATLAS \citep{claret2011} stellar atmosphere models. The \exofastv implementation assumes a Gaussian prior on the limb darkening coefficients with a mean value based on the the Claret table entry that most closely matches the \teff, $\log{g_{*}}$, and [Fe/H] of the host star and uncertainties taken from \citet{claret_bloemen}. However, the dependence of the Claret tables on properties like \teff, $\log{g_{*}}$, and [Fe/H] can introduce model-dependent systematic uncertainties. With a large samples of high cadence transit observations from \tess, we can now attempt directly fitting limb darkening coefficients based on light curves. The estimates of limb darkening parameters themselves can alter estimates of parameters like the duration of the transit using the full width at half maximum depth ($T$) or the duration of planetary ingress or egress ($\tau$), which in turn provide constraints on properties such as $\log(g_{p})$ (Eq. \ref{eqn:gpobservables}) or $\rho_{*}$ (Eq. \ref{eqn:rhostar}). Therefore, it is important to assess the magnitude of the systematic uncertainty due to the adoption of priors on the limb-darkening on these derived properties.

%%%%%%%%%%%%%%%

Finally, we consider the impact of adopting a Gaussian prior on the \teff\ derived from high-resolution spectra. Spectroscopic estimates of \teff\ are commonly available where there exist precise RV follow-up, but are not available for all systems. In those instances, the only direct constraints in \teff\ come from the SED. A spectroscopic prior on \teff\ enters into the analysis in several ways. First, it can provide a more precise estimate of \teff\ and thus $R_*$ as determined from fitting the broad-band SED and parallax. Second, it provides an important input parameter that, when combined with stellar evolutionary tracks or the Torres relations, allow for constraints on $M_*$ and $R_*$. Finally, the \teff\ is one of the main parameters that determines the prior value of the limb darkening from the Claret tables. It is therefore of interest to estimate the systematic error induced by adopting a prior on \teff.

%%%%%%%%%%%%%%%%%%%%%%%%%%%%%%%%%%%%%%%%%%%%%%%%%%%%%%%%%%%%%%%%%%
\section{KELT-15 as a Test Case}\label{sec:kelt15b}

We choose to analyze KELT-15 \citep{kelt15b} as it is a relatively standard transiting Hot Jupiter system. Particularly, it has reasonably well-constrained parameters based on multiple ground based transit and radial velocity observations. Select parameters for KELT-15 are shown in Table \ref{tab:system_params}. The temperature of the host star is fairly typical for known Hot Jupiter systems (\teff\ $\sim 6,000K$) and it is relatively bright ($\rm M_{gaia} \sim 11.1$). \citet{kelt15b} found no evidence for a stellar companion. Follow up from \citet{Mugrauer2019}, did find evidence of a very faint $\rm M_{gaia} \sim 17.7$ M-dwarf companion at a separation of 6 arcseconds or $\sim 2000$ AU. The uncertainty in the broadband SED flux falls around the $\sim 2\%$ level. If one were to assume perfect blending with the companion, the companion would contribute $\sim 0.2\%$ of the total flux. This is an order of magnitude smaller than the uncertainties in the broad band photometry.

The mass of the host star KELT-15 is $\sim 1.2 M_\odot$ \citep{kelt15b}. This puts it in a regime where it is reasonable to fit the host star with each of the MIST, YY, Torres relations, and SED stellar characterization methods. Its host star is bright ($V = 11.2$) \citep{hoeg_2000}, which allows for follow up by \tess. The initial discovery paper also provided estimates for $\rm [Fe/H]$ and \teff\ based on the analysis of medium-resolution spectra \citep{kelt15b}. These characteristics are combined with a modest suite of radial velocity observations and extensive photometric follow-up. Such a data set is reminiscent of the state of available observations for many hot Jupiters. Thus, KELT-15b is fairly representative of the modern sample of characterized hot Jupiters.

KELT-15b was discovered by the the KELT (Kilodegree Extremely Little Telescope) survey \citep{kelt,kelt_south}. The KELT survey observations triggered on planet candidates and recommended them for follow up by their network of associated observatories. Therefore, each KELT planet candidate was observed by several ground based observatories through both relative photometry and radial velocity observations. Photometry was obtained from KELT-South \citep{kelt_south}, LCOGT \citep{lco}, PEST Observatory (owned and operated by TG Tan), and Adelaide Observatory (owned and operated by Ivan Curtis). Radial velocity observations were provided by the CORALIE instrument at the ESO La Silla Observatory \citep{coralie} and and the Anglo-Australian Telescope (AAT) using the CYCLOPS2 instrument \citep{cyclops}.

\begin{table*}
	\centering
	\caption{A summary of KELT-15 parameters from the literature.}
	\label{tab:system_params}
	\begin{tabular}{lccc}
        
	\hline
	Parameter & Description & Value & References \\
	\hline
	TYC & 	Tycho-2 Name & 	8146-86-1 & 	\citet{hoeg_2000} \\
	2MASS & 	2MASS Name & 	J07493960-5207136 & 	\citet{cutri_2003} \\
	TIC & 	TESS ID & 	268644785 & 	\citet{Stassun_2019} \\
	$\alpha_{J2000}$ & 	Right Ascension (R.A.) & 	07:49:39.606 & 	\citet{hoeg_2000} \\
	$\delta_{J2000}$ & 	Declination (Decl.) & 	-52:07:13.58 & 	\citet{hoeg_2000} \\
	$\mu_{\alpha}$ & 	Proper Motion in R.A. (mas yr$^{-1}$) & 	$-3.4 \pm 2.3$ & 	\citet{zacharias_2004} \\
	B$_{T}$& 		Tycho B$_{T}$ mag& 	$ 11.889\pm0.084 $ 	 &	\citet{hoeg_2000}\\
	V$_{T}$& 		Tycho V$_{T}$ mag& 	$ 11.44\pm0.089 $ 	 &	\citet{hoeg_2000}\\
	J& 		2MASS J mag& 	$ 10.205\pm0.02 $ 	 &	\citet{cutri_2003}\\
	H& 		2MASS H mag& 	$ 9.919\pm0.02 $ 	 &	\citet{cutri_2003}\\
	K& 		2MASS K mag& 	$ 9.854\pm0.03 $ 	 &	\citet{cutri_2003}\\
	WISE1& 		WISE1 mag& 	$ 9.787\pm0.03 $ 	 &	\citet{cutri_2014}\\
	WISE2& 		WISE2 mag& 	$ 9.817\pm0.03 $ 	 &	\citet{cutri_2014}\\
	WISE3& 		WISE3 mag& 	$ 9.787\pm0.037 $ 	 &	\citet{cutri_2014}\\
	\textit{Gaia}$_{G}$& 		\textit{Gaia} mag& 	$ 11.069\pm0.02 $ 	 &	\citet{gaia_2016}\\
	\textit{Gaia}$_{BP}$& 		\textit{Gaia}$_{BP}$ mag& 	$ 11.372\pm0.02 $ 	 &	\citet{gaia_2016}\\
	\textit{Gaia}$_{RP}$& 		\textit{Gaia}$_{RP}$ mag& 	$ 10.631\pm0.02 $ 	 &	\citet{gaia_2016}\\
	$\mu_{\delta}$ & 	Proper Motion in Decl. (mas yr$^{-1}$)& 	$-2.0 \pm 2.9$ & 	\citet{zacharias_2004} \\
	GAIA DR2 Distance & 	Estimated Distance (pc)& 	$291 \pm 30$ & 	\citet{kelt15b} \\
        GAIA DR3 Distance & 	Estimated Distance (pc)& 	$314 \pm 2$ & 	\citet{Gaiadr3} \\
	RV & 	Absolute RV (km s$^{-1}$)& 	$7.6 \pm 0.4$ & 	\citet{kelt15b} \\
	$M_{*}$ & 	Mass($M_{\odot}$)& 	$1.181^{+0.051}_{-0.050}$ & 	\citet{kelt15b} \\
	$R_{*}$ & 	Radius($R_{\odot}$)& 	$1.481^{+0.091}_{-0.041}$ & 	\citet{kelt15b} \\
	$L_{*}$ & 	Luminosity ($L_{\odot}$)& 	$2.56^{+0.35}_{-0.20}$ & 	\citet{kelt15b} \\
	T$_{eff}$ & 	Effective temperature (K)& 	$6003^{+56}_{-52}$ & 	\citet{kelt15b} \\
	$\rm [Fe/H]$ & 	Metallicity& 	$0.047 \pm 0.032$ & 	\citet{kelt15b} \\
 
        T$_{eff}$ & 	Effective temperature (K)& 	$6428 \pm 72$ & 	\citet{sweetcat} \\
	$\rm [Fe/H]$ & 	Metallicity& 	$0.24 \pm 0.05$ & 	\citet{sweetcat} \\
	\hline
	\end{tabular}
\end{table*}

\citet{kelt15b} determined that KELT-15 is a G0 type star, roughly 4.6 Gyr in age. In Table \ref{tab:system_params} we also show the broad-band absolute photometry we used to construct our SED. These magnitude observations are the same observations used in \citet{kelt15b} aside from the exclusion of the APASS magnitudes used in that study and the addition of the Gaia DR3 magnitude measurements included here. \cite{kelt15b} found that KELT-15b is a hot Jupiter with a radius of $1.443^{+0.110}_{-0.057} R_{J}$ and a roughly 3 day orbit. The stellar effective temperature and metallicity were determined using Spectroscopy Made Easy \citep{SME}. The observations were originally fit using a modified version of the first iteration of EXOFAST \citep{exofast1}. The authors fit the system twice using constraints on the host star properties from the YY model in one fit and the Torres relations in another. They also found the eccentricity of the system to be consistent with zero to within $2\sigma$.

Additionally, due to its 3 days period we assume KELT-15b's orbit is nearly circular \citep{Adams:2006,Goldreich:1966}. Using Equation 3 in \citet{Adams:2006}, we can roughly expect a circularization timescale of 820 Myr, far shorter than age of system of 4.6 Gyr as estimated by \citet{kelt15b}.

In our re-analysis of KELT-15b we include recent \tess\ observations which were not available at the time of the discovery paper. \tess\ provides high-quality relative photometry dataset that can be used to improve the precision of the system parameters.  \tess\ recorded two sets of observations in January and February of 2019 and three sets of observations from the early months of 2021. KELT-15b provides an excellent example of a hot Jupiter with significant follow up observations. These \tess\ data provide tighter constraints on the directly observable light curve properties like $T$, $\tau$, and $\delta$ than were previously possible. Also when appropriate we adopt the corrected \textit{Gaia} DR3 parallax \citep{GaiaeDR3,Lindegren2021}. In 2018, the host star, KELT-15, was also part of a large scale homogeneous analysis of exoplanet host stars. \citet{sweetcat} provided additional high-resolution spectra observations and produced updated estimates of \teff\ and [Fe/H]. The updated \teff\ estimate of 6428 $\pm$ 72 K disagrees with the \citet{kelt15b} of 6003 $\pm$ 56 by 6 sigma. We adopt the updated \citet{sweetcat} values for this analysis with a 2\% uncertainty as recommended by \citet{tayar2022}.

\begin{figure*}
    \includegraphics[width=8.5cm]{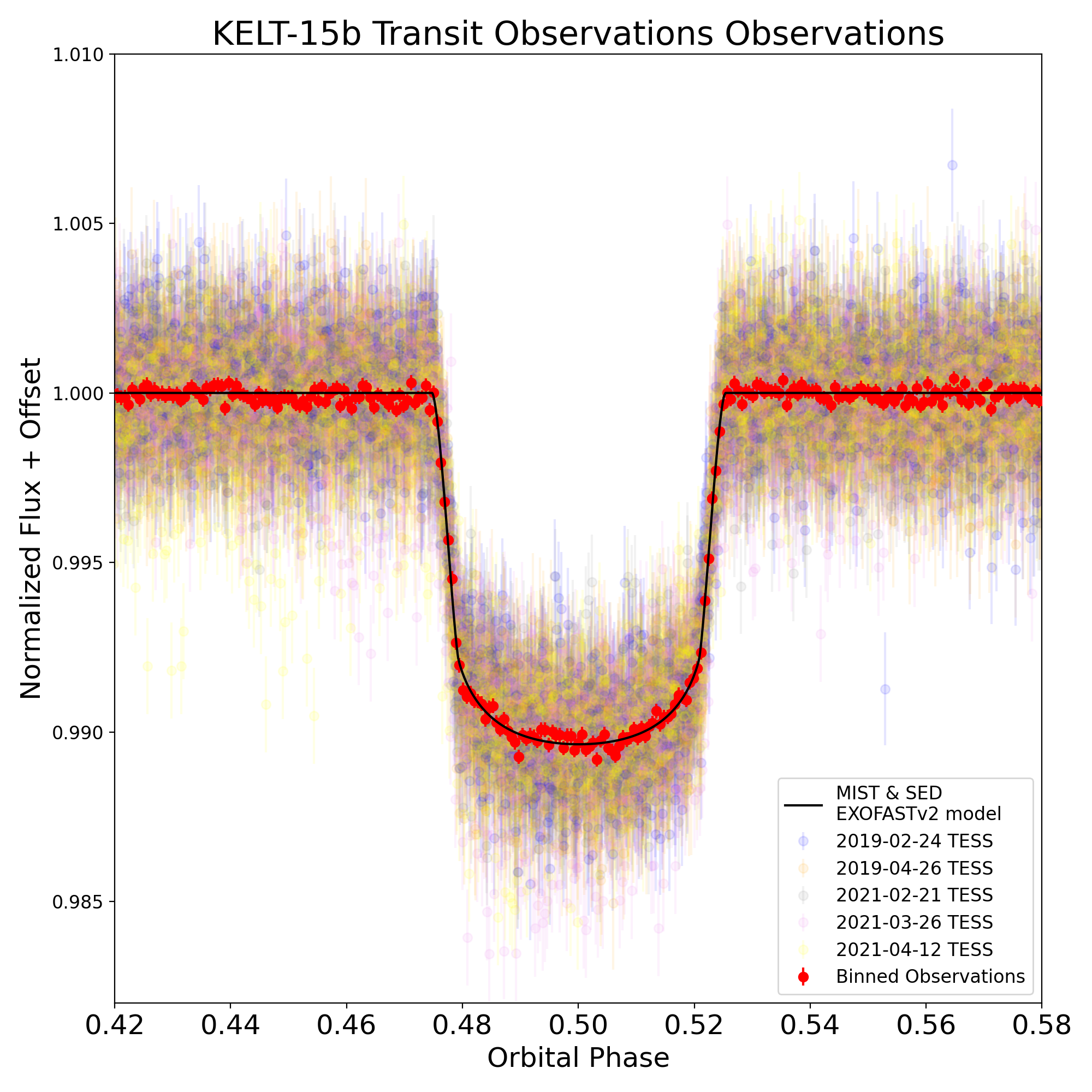}
    \includegraphics[width=9cm]{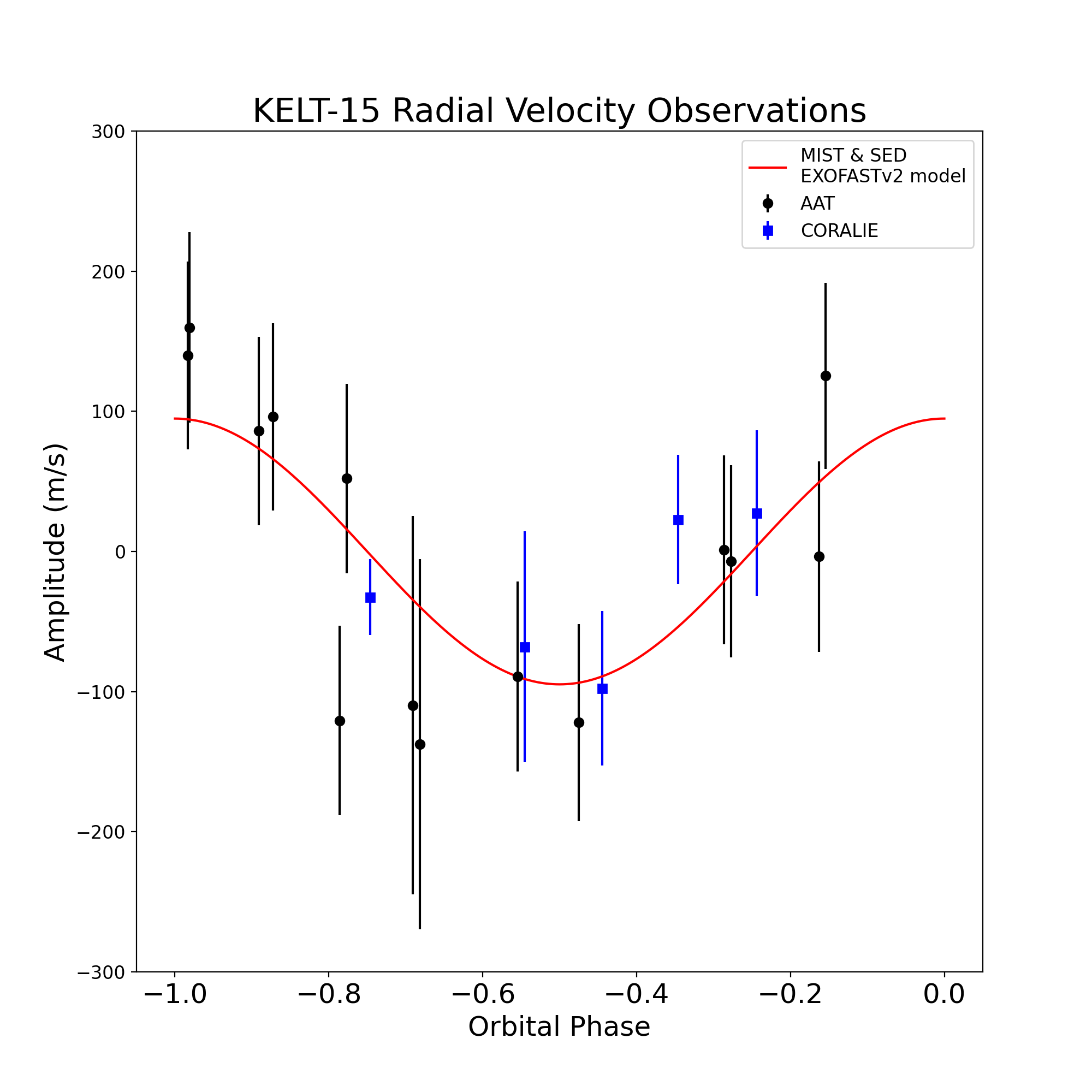}
    \caption{(Left) TESS relatively photometry phase-folded about the best-fit period and centered around the transit. The black line represent the \exofastv model fitting the system with the MIST stellar constraints. The colorful photometric points correspond with individual TESS campaigns. The red points represent all observations binned to 50 points per bin for visualization purposes. (Right) Radial velocity observations of KELT-15 phase-folded about the best-fit period.  These data are from the \citet{kelt15b} discovery paper.}
    \label{fig:obs}
\end{figure*}

The time series TESS transit observations used in this study were downloaded from the Mikulski Archive for Space Telescopes (MAST). We de-trended the SAPFLUX using the {\tt wotan} package, masking transits based on the published period, as well as the transit width $T$ and  the central transit time $t_0$, as estimated by eye \citep{wotan}. This process removes stellar noise and other trends. We ensure that both the transit observations and the radial velocity report their timestamps in $\rm BJD_{TDB}$ \citep{eastman_2010} and that the radial velocity measurements are reported in m/s as required by {\tt EXOFASTv2}. We show the phase folded transit and radial observations used in this study in Figure \ref{fig:obs}.

%%%%%%%%%%%%%%%%%%%%%%%%%%%%%%%%%%%%%%%%%%%%%%%%%%%%%%%%%%%%%%%%%%
\section{Overview of the Analysis} \label{sec:analysis}

\subsection{Default Model}\label{sec:Default}

We consider the ``default" model to be one where only an individual method of breaking the mass-radius degeneracy in the host star is used, and the previously-discussed priors are adopted (Claret LD prior, spectroscopy \teff\ prior, and a circular orbit). The individual stellar characterization techniques we consider are the YY, MIST, Torres, and SED+Gaia constraints as discussed in \S \ref{sec:StellarModels}.

Therefore, the ``default" MIST model uses only the MIST isochrones to break the mass-radius degeneracy and applies all three priors, whereas the ``default" Torres model we would use only the Torres relations and apply the three priors. Using more than a single constraint or change in the priors is no longer considered a ``default" model. By employing only one method of breaking the mass-radius degeneracy at a time, we can compare each method to the others as they use roughly independent constraints. This also allows us to investigate the results of the constituent parts before multiple stellar characterization techniques are applied together in the combination models.

Once the observations from \tess\ are reduced, we follow a set procedure for each iteration. Using {\tt EXOFASTv2}, we provide Gaussian priors on \teff\ and [Fe/H] with medians based on the high-resolution spectra of KELT-15, using the updated values from \citet{sweetcat}. We provide a Gaussian prior on the parallax from \textit{Gaia} Data Release 3 \citep{GaiaeDR3,gaia_2016} with the \citet{Lindegren2021} corrections. Finally, we provide Gaussian priors on the limb darkening coefficients based on the Claret tables \citep{claret2017} and fix eccentricity at zero.

We then model the entire system using all detrended transit and radial velocity observations. We begin by applying only one degeneracy breaking method at time, modelling both the planet and the star. Next, we test for convergence using the Gelman-Rubin statistic \citep{GelmanRubin} as described in \S \ref{sec:Exofast}. We adopt the threshold of a Gelman-Rubin statistic of $<$ 1.01 to indicate convergence, as suggested by \citet{exofastv2}. If the fit is not converged (eg. Gelman-Rubin $> 1.01$) we use that iteration to inform the next run by providing starting values for each parameter. We continue to run iterations of the fit using \exofastv until a Gelman Rubin score of 1.01 is achieved and the fit is considered converged.

%%%%%%%%%%%%%%%%%%%%%%%%%%%%%%%%%%%%%%%%%%%%%%%%%%%%%%%%%%%%%%%%%%
\section{Validation of Fitting Technique}\label{sec:SanityCheck} 

We validating our procedure by using \exofastv to replicate the results in the discovery paper \citep{kelt15b}. Specifically, we use the same observations, \teff\ and [Fe/H] values, and assign the same Gaussian priors used by \citet{kelt15b} to infer the parameters of the system. Thus, in principle, we would expect that the inferred parameters in our analysis would agree to $\sim 0.05\sigma$, where $\sigma$ is the uncertainty on a parameter, due solely to numerical noise arising from finite sampling of the likelihood surface (see \S 23.4 of \citealt{exofastv2}).  Replicating the KELT-15b discovery work therefore provides a `sanity check' to ensure that our analysis methodology is correct. However, we note that the initial analysis by \citet{kelt15b} was conducted using \multifast\, an intermediate upgrade to the original \exofast package \citep{exofast1}.

We would expect greater differences when comparing parameters estimated by different software packages due to subtle differences in model fitting. These differences could be in the fixing limb darkening coefficients to reference table values or allowing interpolation between reference values. Different methods of sampling a parameter space, such as sampling in a linear space or in a log space, can also affect the posterior distribution of parameters. For example, the work of \citet{Gilbert2022} show that the method of ``umbrella sampling" can provide improved uncertainties in properties like the impact parameter or eccentricity compared to the typical direct sampling by more fully exploring extreme regimes. While we do expect differences larger than $0.05\sigma$ between estimates produced by \multifast and {\tt EXOFASTv2}, we still expect close agreement due to their shared basis in the original {\tt EXOFAST} software package. Accordingly, we find subtle differences in some of the inferred parameters as shown in Table \ref{tab:Sanity}. In this replication analysis, we use the same Gaussian priors on \teff, [Fe/H], and the period of the planet as estimated from KELT observations as were used in  the original \citet{kelt15b} analysis.

We compare this initial ``default" run to the case in \citet{kelt15b} which similarly assumes circular orbits and adopts the Torres relations as the mass-radius degeneracy breaking method \citep{Torres_2009}. The results of this trial are shown in Table \ref{tab:Sanity}. In the last column, $\sigma$ is defined as the difference between our median value and the median value of \citet{kelt15b} divided by the larger of the two uncertainties.

\begin{table*}
	\centering
	\caption{We present the results of a re-analysis of KELT-15b using the ground-based observations presented in \citet{kelt15b} while imposing a circular orbit and applying the Torres relations. We compare this initial ``default" run to the case in \citet{kelt15b} which similarly assumes circular orbits and adopts the Torres relations as the mass-radius degeneracy breaking method \citep{Torres_2009}. In the last column, $\sigma$ is defined as the difference between our median value and the median value of \citet{kelt15b} divided by the larger of the two uncertainties. }
	\label{tab:Sanity}
	\begin{tabular}{lccrc}
	\toprule	& Rodriguez et al. & This Work & $\sigma$ & Units	 \\
	\toprule
	$M_*$& $ 1.216^{+0.057}_{-0.057} $& $ 1.220^{+0.060}_{-0.056} $& $-0.07$& $(\msun)$	\\
	$R_*$& $ 1.493^{+0.082}_{-0.082} $& $ 1.502^{+0.092}_{-0.045} $& $-0.11$& $(\rsun)$	\\
	$L_*$& $ 2.65^{+0.32}_{-0.32} $& $ 2.69^{+0.35}_{-0.21} $& $-0.13$& $(\lsun)$	\\
	$\rho_*$& $ 0.518^{+0.032}_{-0.032} $& $ 0.510^{+0.034}_{-0.077} $& $0.14$& (cgs)	\\
	$\log{g}$& $ 4.170^{+0.040}_{-0.040} $& $ 4.171^{+0.020}_{-0.044} $& $-0.03$& (cgs)	\\
	$T_{\rm eff}$& $ 6021.0^{+60.0}_{-60.0} $& $ 6021.0^{+61.0}_{-61.0} $& $0.00$& (K)	\\
	$[{\rm Fe/H}]$& $ 0.051^{+0.034}_{-0.034} $& $ 0.05^{+0.03}_{-0.03} $& $0.03$& (dex)	\\
	$P$& $ 3.329441^{+1.6e-05}_{-1.6e-05} $& $ 3.329441^{+1.6e-05}_{-1.6e-05} $& $0.00$& Period (days)	\\
	$R_P$& $ 1.453^{+0.098}_{-0.098} $& $ 1.468^{+0.110}_{-0.061} $& $-0.15$& $(\rj)$	\\
	$M_P$& $ 0.93^{+0.22}_{-0.22} $& $ 0.93^{+0.26}_{-0.32} $& $0.00$& $(\mj)$	\\
	$a$& $ 0.04657^{+0.00072}_{-0.00072} $& $ 0.04664^{+0.00075}_{-0.00073} $& $-0.09$& Semi-major axis (AU)	\\
	$i$& $ 88.4^{+1.1}_{-1.1} $& $ 88.3^{+1.2}_{-1.7} $& $0.10$& Inclination (Degrees)	\\
	$T_{eq}$& $ 1645.0^{+41.0}_{-41.0} $& $ 1649.0^{+45.0}_{-26.0} $& $-0.10$& (K)	\\
	$K$& $ 110.0^{+26.0}_{-26.0} $& $ 110.0^{+31.0}_{-38.0} $& $0.00$& (m/s)	\\
	$R_P/R_*$& $ 0.1001^{+0.0021}_{-0.0021} $& $ 0.1005^{+0.0023}_{-0.0021} $& $-0.18$& Radius of planet in stellar radii	\\
	$a/R_*$& $ 6.72^{+0.13}_{-0.13} $& $ 6.69^{+0.14}_{-0.35} $& $0.12$& Semi-major axis in stellar radii	\\
	$\delta$& $ 0.01001^{+0.0004}_{-0.0004} $& $ 0.01011^{+0.00046}_{-0.00043} $& $-0.22$& Transit depth (fraction)	\\
	$\tau$& $ 0.01627^{+0.0019}_{-0.0019} $& $ 0.01644^{+0.00220}_{-0.00079} $& $-0.09$& ingress/egress transit duration (days)	\\
	$T_{14}$& $ 0.1717^{+0.0024}_{-0.0024} $& $ 0.1724^{+0.0026}_{-0.0021} $& $-0.29$& Total transit duration (days)	\\
	$T$& $ 0.1551^{+0.0015}_{-0.0015} $& $ 0.1556^{+0.0017}_{-0.0016} $& $-0.30$& FWHM transit duration (days)	\\
	$b$& $ 0.19^{+0.17}_{-0.17} $& $ 0.20^{+0.18}_{-0.14} $& $-0.10$& Transit Impact parameter	\\
	$\rho_P$& $ 0.363^{+0.11}_{-0.11} $& $ 0.35^{+0.12}_{-0.14} $& $0.10$& Density (cgs)	\\
	$logg_P$& $ 3.03^{+0.10}_{-0.10} $& $ 3.01^{+0.12}_{-0.20} $& $0.13$& Surface gravity	\\
	\hline
	\end{tabular}
\end{table*}

Overall, we find agreement to within $0.3\sigma$ for all parameters, but note that some parameters are more discrepant than others. Surprisingly, the parameters with the largest discrepancies are the (nearly) direct observables of the transit depth and the transit duration using the full width at half the maximum depth $T$. As shown in Table \ref{tab:Sanity}, the median values of these parameters differ by $0.2-0.3\sigma$. The estimated values of these 'direct observables' should be independent of most of the priors, with the exception of those for the limb darkening. We suspect that the most likely cause of the discrepancies in our estimated parameters is the different approaches of the two versions of EXOFAST in regards to limb darkening. The original version of EXOFAST and \multifast directly adopted the limb darkening coefficients from the Claret 2011 tables \citep{claret2011}. \exofastv now includes the Claret 2017 tables \citep{claret2017} and applies Gaussian priors based on the limb darkening coefficients values rather than fixing them at Claret table values.

Agreement to better than $0.3\sigma$ does instill confidence that our fitting procedure is working properly. However, given the convergence level required by \exofastv we would expect agreement between the parameters on the level of $0.05\sigma$ \citep{exofastv2}. Even when using fitting packages with the same heritage, these differences provides motivation for fitting a large sample of exoplanets with a consistent and uniform methodology.

We expect the differences between the stellar and exoplanetary parameters inferred by \exofastv and by software packages with other heritage to be even larger. For example, the work of \citet{patel2022} showed that limb darkening coefficients derived from model atmospheres can have a differences in $u_{2}$ as large as $\sim 0.2$ compared to empirically obtained limb darkening coefficients from TESS observations. If other packages use differing approaches to fitting limb darkening coefficients or to sampling parameter spaces, we would expect corresponding differences in the characterization of the transiting exoplanet.

\subsection{Linear vs Log K}

\begin{figure*}
    \centering
    \includegraphics[width=8.5cm]{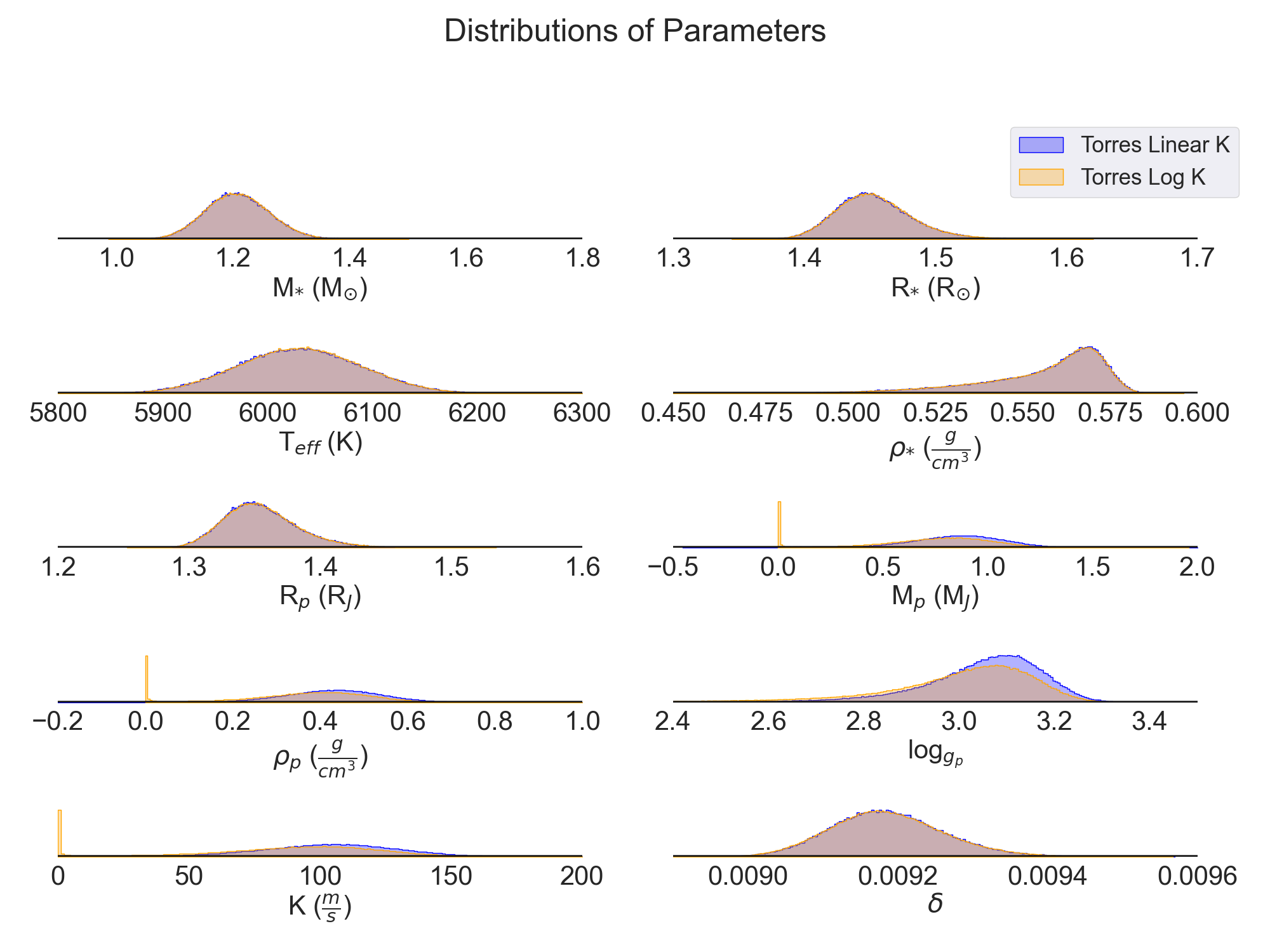}
    \includegraphics[width=8.5cm]{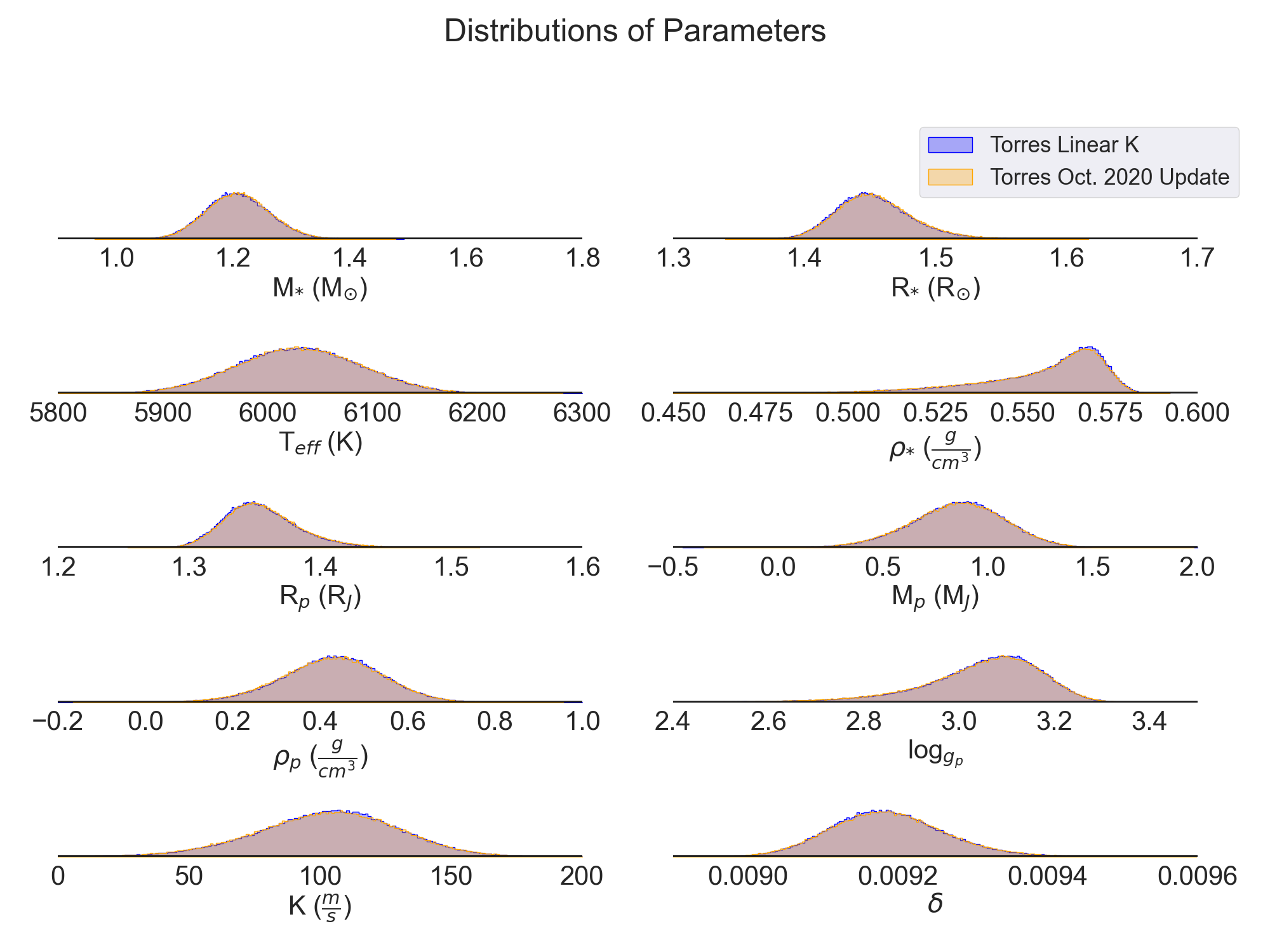}

    \caption{The overlapping histograms show the posterior distributions of several directly observable ($K$, $\delta$,$\rho_{*}$, $\log{g_{p}}$) and derived  ($M_{*}$, $R_{*}$, \teff\, $M_{p}$, $R_{p}$, $\log{g_p}$ ) parameters in the KELT-15 system. Both models use the Torres relations as constraints on the host star. In the left panel, the blue histograms show the posteriors assuming a uniform prior in $K$, where as the yellow histograms show the posterior distributions assuming a uniform prior in $\log{K}$. In the right panel, we compare the posteriors assuming a uniform prior in $K$ to the posteriors assuming a uniform prior in planet mass, as implemented in the \exofastv October 2020 update. We see good agreement between these methods and no longer see the posterior distribution build up around $K\sim 0$.}\label{fig:linlogK}
\end{figure*}

%Made with LargeHistogram_v5.py

After validating our methodology with the same data set as \citet{kelt15b}, we proceed to include TESS observations while still maintaining the \teff\ and [Fe/H] values from \citet{kelt15b}. We first fit the light curves (including TESS) and radial velocity data using the stellar constraints from the Torres relations.  We found small but non-negligible support for $K\sim0$ in the form of a build-up in the posterior at low $K$ (see Figure \ref{fig:linlogK}). At that time, \exofastv stepped uniformly in $\log{K}$, and thus effectively adopted a uniform distribution in $\log K$ rather then a uniform distribution in $K$.  Since the former prior has more support for lower values of $K$, this produces a build-up near $K\sim0$ when the RV data are insufficiently precise.  All quantities that are derived from $K$, such as the planetary radius and density, will also display some support for values near 0.  

We note that we expected to see this behavior in our previous fits to the KELT-15b datasets without TESS, but did not. This is surprising because the addition of the TESS data should not impact the RV fit. We suspect that the improved precision in the ephemeris of KELT-15b (period and time of transit) subtly impacts the fit to the RV data.

\exofastv also provides the option of adopting a linear prior in $K$.  Not surprisingly, in this case the posterior does not show a build-up near $K\sim0$, and the median value of $K$ is much closer to that reported by \citet{kelt15b}. This result indicates that a minor change in the method of sampling a parameter space can result in a significantly altered median value. This is especially apparent in parameters that are not well constrained, like this instance of $K$. When sampling $\log{K}$ space we find a median planet mass of $0.76^{+0.25}_{-0.38} M_{J}$. In contrast, when sampling $K$ space we find a median of $0.87^{+0.22}_{-0.24} M_{J}$. This is a fractional difference of approximately 13.5\% or $0.3\sigma$, assuming the uniform $K$ space sampling version as the fiducial. 

This is a $0.3\sigma$ difference which is solely derived from changing the form of a prior within one planet modeling package. Although this change is less than $1\sigma$, this underlines the need for consistent methodology in parameter fitting. Understanding these systematic uncertainties is necessary to robustly compare the derived parameters of different exoplanet systems. Of course, this chosen method of sampling $K$ space will have a smaller effect for systems with radial velocity data sets with higher statistical power.

Later updates in October of 2020 to EXOFASTv2 changed from a uniform prior in $\log K$ to a uniform prior in the planet mass $M_p$. As Figure \ref{fig:linlogK} shows, the results derived from applying a uniform prior on $K$ and applying a uniform prion on $M_p$ agree extremely well, as expected. This method of stepping in planet mass space resolves the build up around $K\sim 0$ and produces a median planet mass of $0.86^{+0.22}_{-0.24} M_{J}$. For the remainder of this paper, our default method will adopt a uniform prior in $M_p$.

%%%%%%%%%%%%%%%%%%%%%%%%%%%%%%%%%%%%%%%%%%%%%%%%%%%%%%%%%%%%%%%%%%
\section{Results} \label{sec:results}

We now re-analyze KELT-15b while incorporating space based transit photometry from \tess\ and the updated estimates of \teff\ and [Fe/H] from \citet{sweetcat} with a 2\% error floor in \teff\ as recommended by \citet{tayar2022}. We present the posterior distributions from fitting the system with \exofastv as described in \S \ref{sec:analysis}. We vary the method of characterizing the host star as well as changing several underlying assumptions about the system.

\subsection{Results of Default Model}\label{sec:DefaultResults}

%Old version of figure
%\includegraphics[width=14cm]{May23_prelim_paper_partv31.png}

\begin{figure*}
    \centering
    \includegraphics[width=14cm]{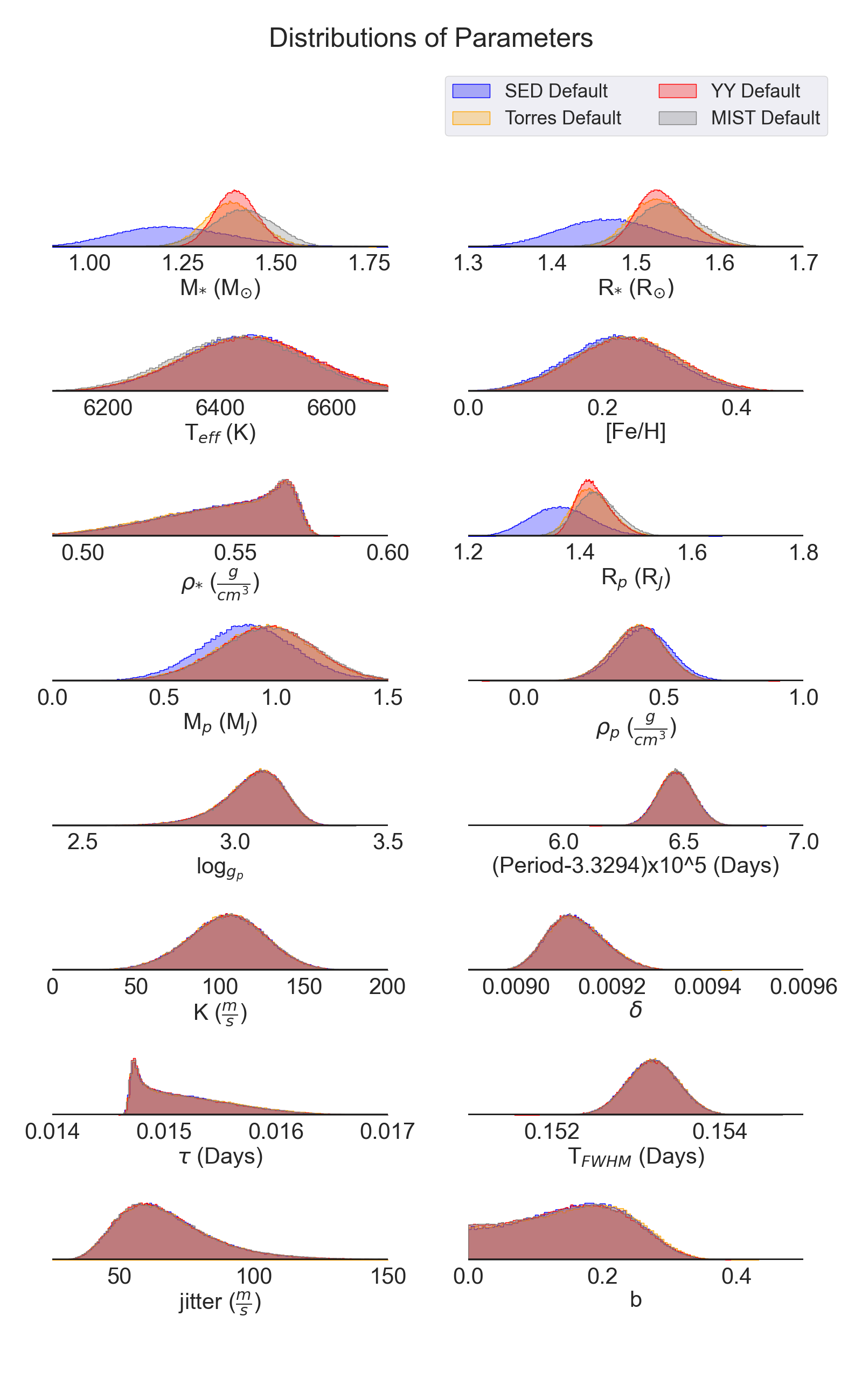}
    
    \caption{The histograms represent the posterior distributions of selected parameters of the KELT-15 system from our ``default" stellar characterization methods: Torres, YY, MIST, and SED. We assume circular orbits, include a Gaussian prior on \teff\ from high-resolution spectroscopy \citep{sweetcat}, and use constraints from the Claret limb darkening tables \citep{claret2017}.}\label{fig:DefaultResults}
\end{figure*}

Figure \ref{fig:DefaultResults} shows the posterior distributions for several system parameters for all four single-constraint models (Torres, YY, SED, MIST). These include the ``default assumptions" outlined in \S \ref{sec:analysis} of a circular orbit, a prior on \teff\ based on \citet{sweetcat}, and priors on the limb darkening coefficients based on the Claret limb darkening tables. The shaded regions in Figure \ref{fig:DefaultResults} represent histograms of the posterior distributions for each parameter as calculated by \exofastv for each single-constraint model.

We see that all four single-constraint models (Torres, YY, SED, MIST) closely agree with each other regarding the directly observable parameters ($P$, $K$, $\delta$, $\tau$, $\log{g_{p}}$, $T$, $\rho_{*}$, and RV jitter). The posterior distributions of these parameters overlap almost exactly, and are insulated from any effect of changing the stellar modeling method used in our ``default" method.  This is because none of the individual constraints independently constrain $\rho_*$, and thus do not provide any constraints on these directly-observable parameters. When compared to the original discovery paper, we find a significant decrease in the uncertainty in the period by 50\%, due to the inclusion of \tess\ observations, which provide a longer baseline. 

We see similar reductions in uncertainty for the derived parameters ($\rm M_{*}$, $\rm R_{*}$, Stellar Age, $\rm R_{p}$, $\rm M_{p}$, etc.). We find a fractional decrease in the uncertainty in the planetary radius of $\sim 65\%$ between the YY-only case in \cite{kelt15b} and the YY-only case of this work. This is primarily due to a 42\% decrease in the uncertainty in the stellar radius. Such a decrease in the uncertainty of the stellar radius is likely due to the additional photometric transit observations which provide improved constraints on the stellar density. \tess\ observations provide useful constraints on the density of the host star and the transit depth with fractional improvements of 84\% and 87\% in the uncertainties of these parameters respectively when compared to the \citet{kelt15b} study. We see that all of the single constraint methods we consider, including the \citet{kelt15b} values, result in a consistent stellar density. Table \ref{tab:result_params} presents our findings for select system parameters. Our complete suite of results for the single constraint default fits are shown in Table \ref{tab:CanonicalTable}.

We find similar consistency amongst the physical parameters as noted in the observable parameters. Notably, we see close agreement between the Torres, YY, and MIST methods. In this case we find a fractional difference between the Torres and MIST estimates of $\rm R_{*}$ of 0.6 \% or $0.2 \sigma$, where we define $\sigma$ to be the difference in the medians divided by the larger of the two uncertainties.

These approaches agree in $\rm R_*$ to the $1.1 \sigma$ level and in $\rm M_{*}$ to the $1.2 \sigma$ level when comparing the most discrepant estimate from the MIST models and the SED approach. Estimates of \teff\ agree within $0.2 \sigma$, showing good agreement between the spectroscopic prior from \citet{sweetcat} and the SED information. We see this minor discrepancy in $R_*$ directly translated to the estimate of planetary radius at the $1.1 \sigma$ level. In the estimate of planetary mass however, the discrepancy in the host star mass estimate is muted ($0.4 \sigma$) due to the larger uncertainty in $K_*$. However, these discrepancies are close to or below $1 \sigma$ showing strong agreement between all four methods of host star characterization. The uncertainty introduced by changing the method of breaking the stellar mass-radius degeneracy is of similar magnitude to the statistical uncertainty in all the primary physical parameters ($\rm M_{*}$, $\rm R_{*}$, $\rm M_{p}$, $\rm R_{p}$, etc.).

\begin{table*}
	\centering
	\caption{We display selected parameter values for our ``default" fits of the KELT-15 system.}
	\label{tab:result_params}
	\begin{tabular}{lcccccl}
	\toprule	& MIST & YY & Torres & SED & Range & Units	 \\
	\toprule
	$M_*$& $ 1.409^{+0.086}_{-0.089} $& $ 1.391^{+0.057}_{-0.056} $& $ 1.382^{+0.074}_{-0.071} $& $ 1.22^{+0.17}_{-0.15} $& $0.189$& $(\msun)$	\\
	$R_*$& $ 1.537^{+0.04}_{-0.038} $& $ 1.53^{+0.033}_{-0.028} $& $ 1.528^{+0.037}_{-0.033} $& $ 1.469^{+0.062}_{-0.059} $& $0.068$& $(\rsun)$	\\
	$\rho_*$& $ 0.55^{+0.016}_{-0.026} $& $ 0.55^{+0.016}_{-0.026} $& $ 0.549^{+0.017}_{-0.027} $& $ 0.549^{+0.017}_{-0.026} $& $0.001$& (cgs)	\\
	$\log{g_{*}}$& $ 4.214^{+0.012}_{-0.015} $& $ 4.2124^{+0.01}_{-0.014} $& $ 4.21^{+0.012}_{-0.015} $& $ 4.193^{+0.023}_{-0.026} $& $0.021$& (cgs)	\\
	$T_{\rm eff}$& $ 6430.0^{+120.0}_{-120.0} $& $ 6450.0^{+120.0}_{-120.0} $& $ 6450.0^{+120.0}_{-120.0} $& $ 6450.0^{+120.0}_{-120.0} $& $20$& (K)	\\
	$[{\rm Fe/H}]$& $ 0.233^{+0.079}_{-0.08} $& $ 0.236^{+0.08}_{-0.08} $& $ 0.236^{+0.079}_{-0.079} $& $ 0.225^{+0.078}_{-0.078} $& $0.011$& (dex)	\\
	$\rho_p$& $ 0.409^{+0.094}_{-0.097} $& $ 0.411^{+0.093}_{-0.097} $& $ 0.408^{+0.093}_{-0.096} $& $ 0.426^{+0.098}_{-0.1} $& $0.018$& (cgs)	\\
	$R_p$& $ 1.428^{+0.041}_{-0.037} $& $ 1.421^{+0.035}_{-0.027} $& $ 1.42^{+0.038}_{-0.033} $& $ 1.365^{+0.058}_{-0.055} $& $0.063$& $(\rj)$	\\
	$M_p$& $ 0.96^{+0.22}_{-0.23} $& $ 0.96^{+0.21}_{-0.22} $& $ 0.95^{+0.21}_{-0.22} $& $ 0.87^{+0.21}_{-0.21} $& $0.09$& $(\mj)$	\\
	Period& $ 3.32946468^{+7.5e-07}_{-7.5e-07} $& $ 3.32946467^{+7.7e-07}_{-7.5e-07} $& $ 3.32946466^{+7.6e-07}_{-7.4e-07} $& $ 3.32946467^{+7.6e-07}_{-7.6e-07} $& $0$& Days	\\
	b& $ 0.155^{+0.085}_{-0.099} $& $ 0.155^{+0.085}_{-0.098} $& $ 0.161^{+0.084}_{-0.1} $& $ 0.159^{+0.082}_{-0.097} $& $0.006$& -	\\
	$\log{g_{p}}$& $ 3.068^{+0.087}_{-0.12} $& $ 3.068^{+0.087}_{-0.12} $& $ 3.065^{+0.087}_{-0.12} $& $ 3.066^{+0.088}_{-0.12} $& $0.003$& (cgs)	\\
	K& $ 104.0^{+23.0}_{-24.0} $& $ 104.0^{+23.0}_{-24.0} $& $ 104.0^{+23.0}_{-24.0} $& $ 104.0^{+23.0}_{-24.0} $& $0$& (m/s)	\\
	$\delta$& $ 0.00912^{+6.6e-05}_{-5.5e-05} $& $ 0.009121^{+6.6e-05}_{-5.5e-05} $& $ 0.009124^{+6.7e-05}_{-5.7e-05} $& $ 0.009123^{+6.5e-05}_{-5.5e-05} $& $0$& -	\\
	$\tau$& $ 0.0151^{+0.00058}_{-0.00034} $& $ 0.0151^{+0.00058}_{-0.00034} $& $ 0.01513^{+0.00059}_{-0.00036} $& $ 0.01512^{+0.00057}_{-0.00035} $& $0$& Days	\\
	$T_{14}$& $ 0.16837^{+0.0005}_{-0.00041} $& $ 0.16837^{+0.0005}_{-0.00042} $& $ 0.16839^{+0.00052}_{-0.00042} $& $ 0.16837^{+0.0005}_{-0.00042} $& $0$& Days	\\
	$cos(i)$& $ 0.023^{+0.013}_{-0.014} $& $ 0.023^{+0.013}_{-0.014} $& $ 0.023^{+0.013}_{-0.015} $& $ 0.023^{+0.013}_{-0.014} $& $0$& -	\\
	\hline
	\end{tabular}
\end{table*}

\subsection{Results of Combining Methods of Stellar Fitting}\label{sec:ComboResults}

\begin{figure*}
    \centering
    \includegraphics[width=14cm]{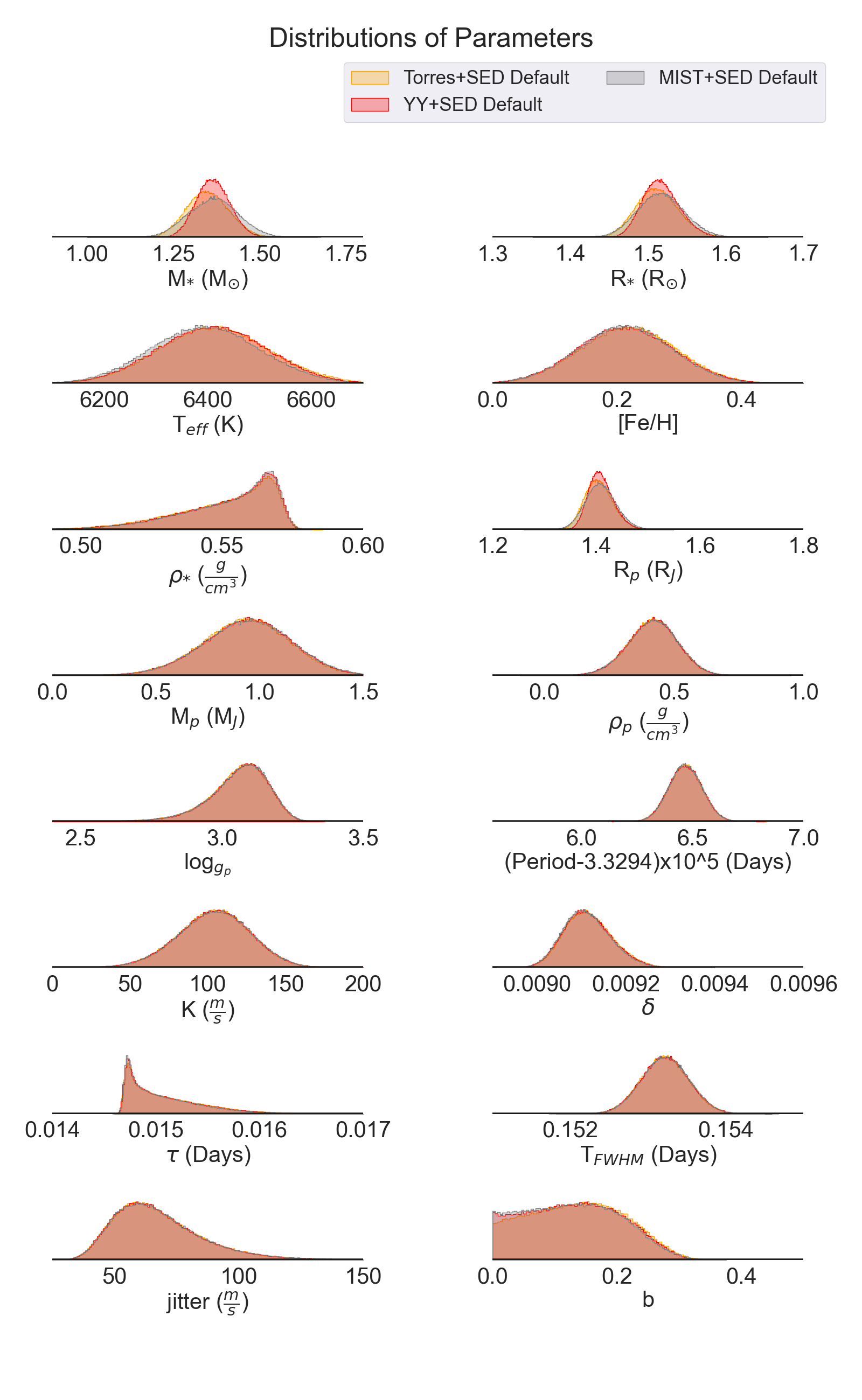}
    
    \caption{We present the results of our combination stellar constraint models: Torres + SED, YY + SED, MIST + SED. We assume circular orbits, adopt a spectroscopically-derived prior on \teff\ from \citet{sweetcat}, and a prior on the limb darkening using the Claret limb darkening tables \citep{claret2017}.  }\label{fig:CombineResults}
\end{figure*}

We next compare the results of combining two methods of breaking the stellar mass-radius degeneracy together. We investigate combining the YY isochrones and the SED constraints, MIST isochrones and the SED, and finally the Torres relations with the SED.  The results of the characterization of the KELT-15 system with these combination methods are shown in Figure \ref{fig:CombineResults} and in Table \ref{tab:result_paramsSED}. The three combination constraint models agree with each other to 0.3$\sigma$ across all parameters. We see similarly tight agreement in the stellar mass estimate at the $ 0.24 \sigma$ level between the MIST+SED and Torres+SED approaches. We see that the median values of the temperature estimate are nearly identical ($0.1 \sigma$).

We find that the additional constraints on $R_*$ from the combined SED, Gaia EDR3, and spectroscopic \teff\ prior from \citet{sweetcat} drive the combination models toward broadly consistent posteriors for all of "observable" and physical parameters.  The full results of the combined models are presented in Tables \ref{tab:CanoncialCombo} and \ref{tab:CanoncialComboCont}.

%Canonical plot
%OLD
%Figs/May23_prelim_papercombo_partv31.png

\begin{table*}
	\centering
	\caption{We display selected parameter values for our ``combination" stellar constraint fits of the KELT-15 system.}
	\label{tab:result_paramsSED}
	\begin{tabular}{lccccl}
	\toprule	& MIST +SED & YY + SED & Torres + SED & Range & Units	 \\
	\toprule
	$M_*$& $ 1.366^{+0.072}_{-0.076} $& $ 1.364^{+0.05}_{-0.048} $& $ 1.382^{+0.074}_{-0.071} $& $0.018$& $(\msun)$	\\
	$R_*$& $ 1.516^{+0.031}_{-0.031} $& $ 1.515^{+0.026}_{-0.022} $& $ 1.528^{+0.037}_{-0.033} $& $0.013$& $(\rsun)$	\\
	$\rho_*$& $ 0.556^{+0.011}_{-0.022} $& $ 0.556^{+0.012}_{-0.023} $& $ 0.549^{+0.017}_{-0.027} $& $0.007$& (cgs)	\\
	$\log{g_{*}}$& $ 4.213^{+0.011}_{-0.014} $& $ 4.2126^{+0.0085}_{-0.012} $& $ 4.21^{+0.012}_{-0.015} $& $0.003$& (cgs)	\\
	$T_{\rm eff}$& $ 6400.0^{+110.0}_{-110.0} $& $ 6410.0^{+110.0}_{-110.0} $& $ 6450.0^{+120.0}_{-120.0} $& $50$& (K)	\\
	$[{\rm Fe/H}]$& $ 0.213^{+0.076}_{-0.076} $& $ 0.213^{+0.079}_{-0.077} $& $ 0.236^{+0.079}_{-0.079} $& $0.023$& (dex)	\\
	$\rho_p$& $ 0.42^{+0.095}_{-0.1} $& $ 0.418^{+0.094}_{-0.099} $& $ 0.408^{+0.093}_{-0.096} $& $0.012$& (cgs)	\\
	$R_p$& $ 1.407^{+0.031}_{-0.029} $& $ 1.407^{+0.027}_{-0.022} $& $ 1.42^{+0.038}_{-0.033} $& $0.013$& $(\rj)$	\\
	$M_p$& $ 0.94^{+0.21}_{-0.22} $& $ 0.94^{+0.21}_{-0.22} $& $ 0.95^{+0.21}_{-0.22} $& $0.01$& $(\mj)$	\\
	Period& $ 3.32946467^{+7.6e-07}_{-7.5e-07} $& $ 3.32946467^{+7.7e-07}_{-7.6e-07} $& $ 3.32946466^{+7.6e-07}_{-7.4e-07} $& $0$& Days	\\
	b& $ 0.127^{+0.083}_{-0.084} $& $ 0.13^{+0.083}_{-0.085} $& $ 0.161^{+0.084}_{-0.1} $& $0.034$& -	\\
	$\log{g_{p}}$& $ 3.072^{+0.087}_{-0.12} $& $ 3.071^{+0.087}_{-0.12} $& $ 3.065^{+0.087}_{-0.12} $& $0.007$& (cgs)	\\
	K& $ 104.0^{+23.0}_{-25.0} $& $ 104.0^{+23.0}_{-24.0} $& $ 104.0^{+23.0}_{-24.0} $& $0$& (m/s)	\\
	$\delta$& $ 0.009109^{+5.8e-05}_{-5e-05} $& $ 0.00911^{+5.8e-05}_{-5.1e-05} $& $ 0.009124^{+6.7e-05}_{-5.7e-05} $& $0$& -	\\
	$\tau$& $ 0.01497^{+0.00048}_{-0.00023} $& $ 0.01498^{+0.00048}_{-0.00024} $& $ 0.01513^{+0.00059}_{-0.00036} $& $0$& Days	\\
	$T_{14}$& $ 0.16826^{+0.00043}_{-0.00038} $& $ 0.16827^{+0.00044}_{-0.00038} $& $ 0.16839^{+0.00052}_{-0.00042} $& $0$& Days	\\
	$cos(i)$& $ 0.018^{+0.013}_{-0.012} $& $ 0.019^{+0.013}_{-0.012} $& $ 0.023^{+0.013}_{-0.015} $& $0.005$& -	\\
	\hline
	\end{tabular}
\end{table*}

\begin{figure*}
    \centering
    %Made with ParameterScatterPlots.py
    \includegraphics[width=14cm]{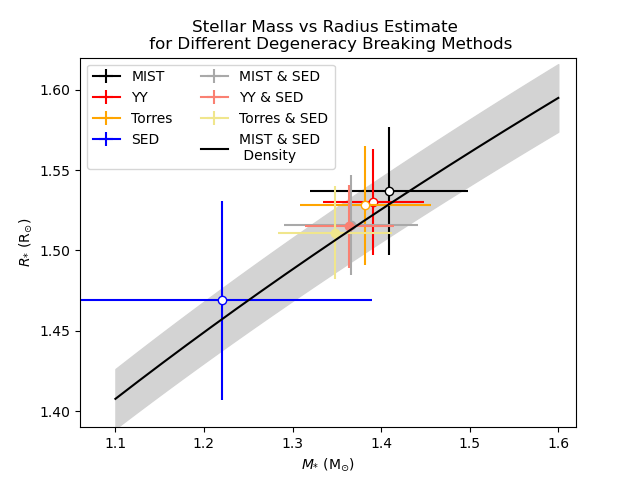}
    \includegraphics[width=14cm]{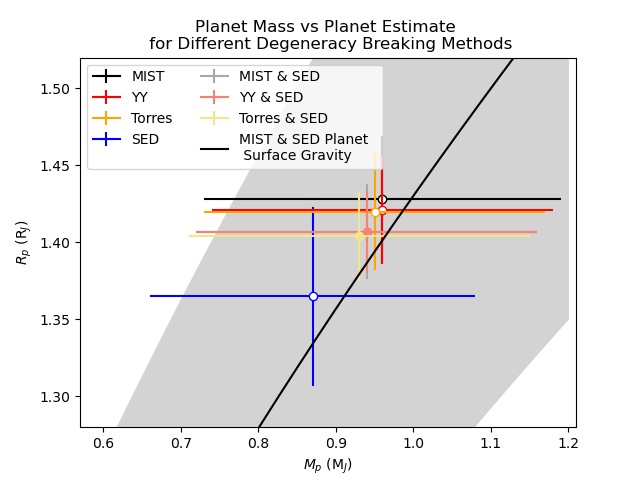}
    \caption{ Mass and radius estimates for the host star are represented in the top panel. Estimates from models using the default and combination stellar constraints are included. The stellar properties are compared to the MIST + SED density estimate represented by the black line with the gray region showing the $1\sigma$ uncertainty. Corresponding estimates for the planet KELT-15b are displayed in the lower panel. The planet properties are compared to the planet surface gravity, inferred from modelling the system with the MIST + SED host star constraints, represented by the black line with the gray region showing the $1\sigma$ uncertainty. The system parameters inferred from single constraints on the host star are shown as open circles, whereas those inferred from combination constraints are shown as filled circles. }\label{fig:DensityAll}
\end{figure*}

To compare these combined methods with the ``default" single-constraint methods, we show the mass and radius estimates in Figure \ref{fig:DensityAll} for both the planet and the host star. We also show the stellar density $\rho_{*}$ as estimated by the MIST + SED fit with the black line with the $1\sigma$ uncertainty shown as the grey band. As expected, since $\rho_*$ is essentially a direct observable, the values $R_*$ and $M_*$ inferred using the various methods are correlated such that they all lie along a track of roughly constant $\rho_*$. Indeed, we find that all methods infer a value of $\rho_*$ that is consistent to within $1\sigma$. As this density estimate is derived largely from the light curve data, this highlights the importance of inferring all of the physical properties by fitting all of the data, including that related to the properties of the star, simultaneously and self-consistently.

%\textcolor{magenta}{we do see that the models are mostly clustered and the SED is still off by itself though, just lower than the models}
%\textcolor{magenta}{Maybe also make a version of the histograms plot for all the no teff models?}

%We find that the SED alone method produces the smallest values for both radius and mass. MIST alone provides the largest estimates. The combination models are consistently located between these two. All of the planet density estimates are extremely consistent with the MIST + SED estimate, primarily due to the large uncertainties in planet mass due to the limited radial velocity observations. We also see that the MIST + SED mass and radius estimates are aligned with the MIST + SED density.

\subsection{Results of Variation in Each Model}\label{sec:VariationResults}

\begin{figure*}
    \includegraphics[width=0.95\textwidth]{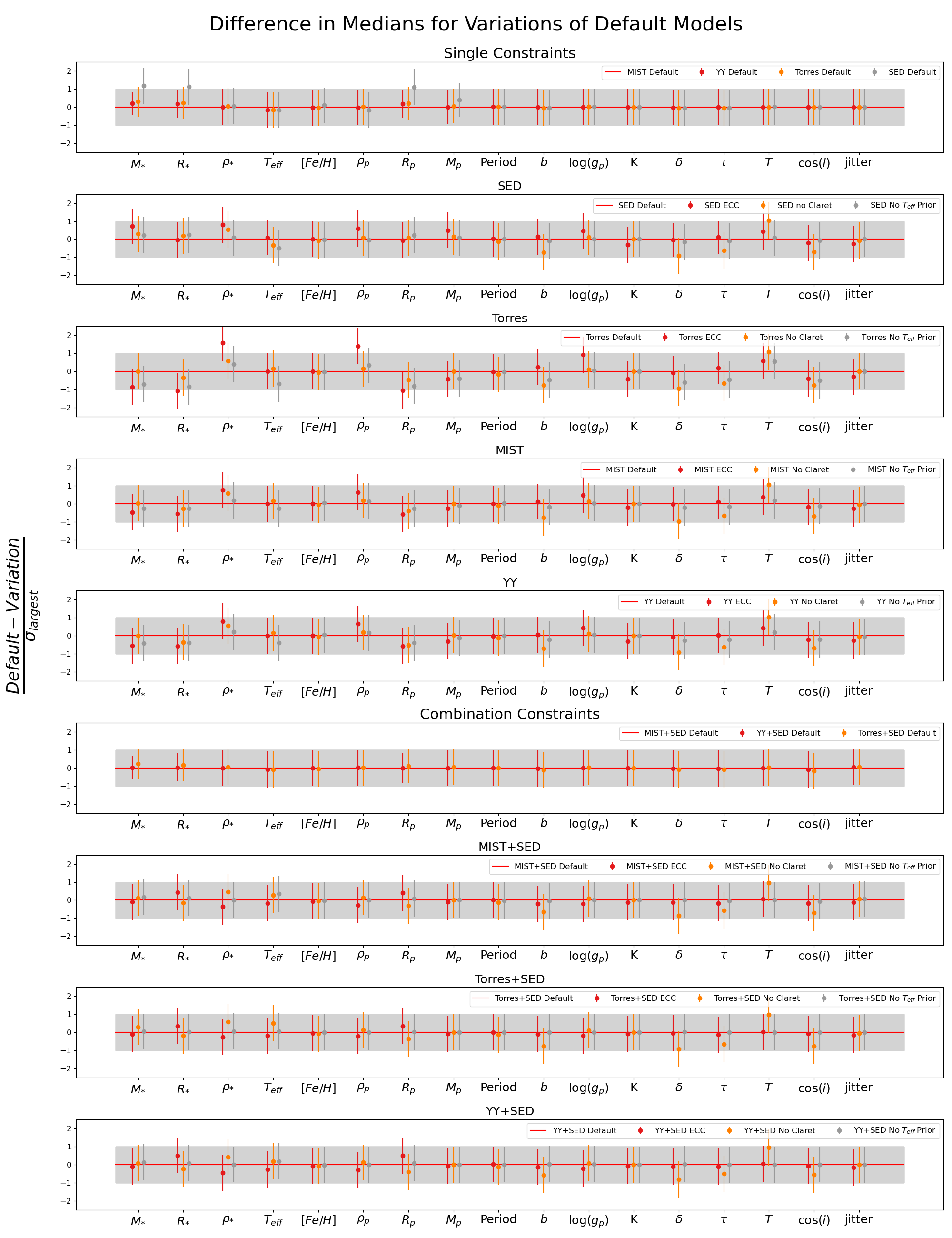}
    \caption{The points with uncertainties show the difference in medians of the variations compared to the ``default" model represented by the red line. We take the difference of the two models and divide by the larger uncertainty to find the $\sigma$ difference represented here. The grey bar represents a $1\sigma$ difference.}\label{fig:VariationMedDiff}
\end{figure*}

We additionally assess the impact of changing three of the priors discussed earlier: 1) the assumption of a circular orbit, 2) adopting a prior on the limb-darkening coefficients using the Claret tables, and 3) adopting a spectroscopically derived prior on \teff. Adopting a different choice for one of these three priors provide a “variation" from one of the seven single and double  constraint models. By changing all three priors in turn we consider a total of 21 variations of the single + double constraint models.  A summary of the results of all of these variations is presented in Figures \ref{fig:VariationMedDiff}, where we show the difference between the “default" model median value and the “variation" model median value for each parameter, normalized by the larger of the statistical uncertainties of each model. The gray bar represents a difference of $1 \sigma$ in this quantity.

%Old Figure
%May23_Comp_Updatedv3.png

A quick perusal of Figure \ref{fig:VariationMedDiff} indicates that inferred medians for nearly all the relevant observable and physical parameters for a 21 different variation models differ from the default values by less than $\sim 1 \sigma$, and in most cases much less than $1 \sigma$.  For the sake of brevity, we will not thoroughly discuss the results of all these variations, but only mention a few highlights.

%\subsubsection{Results of exploring the impact of priors on the limb darkening coefficients}
\subsection{Impact of Adopting a Limb Darkening Prior}\label{sec:VariationClaret}

By default, \exofastv applies a prior on the limb darkening coefficients (LDCs) when fitting the light curve based on the Claret tables \citep{claret_bloemen,claret2017}, where the median value of the prior is determined by the expected LDC for a star with the value of \teff\, [Fe/H], and $log(g)$ for each step in the MCMC chain. Thus the prior on the LDCs influence the fitted value of the limb darkening coefficients as well as the direct observables, which in turn constrain $\rho_{*}$.  Conversely, the light curve weakly constrains \teff, [Fe/H], and $\rm log(g)$ through the information in the LDC in the light curves.  

\begin{figure}
    \centering
    \includegraphics[width=0.5\textwidth]{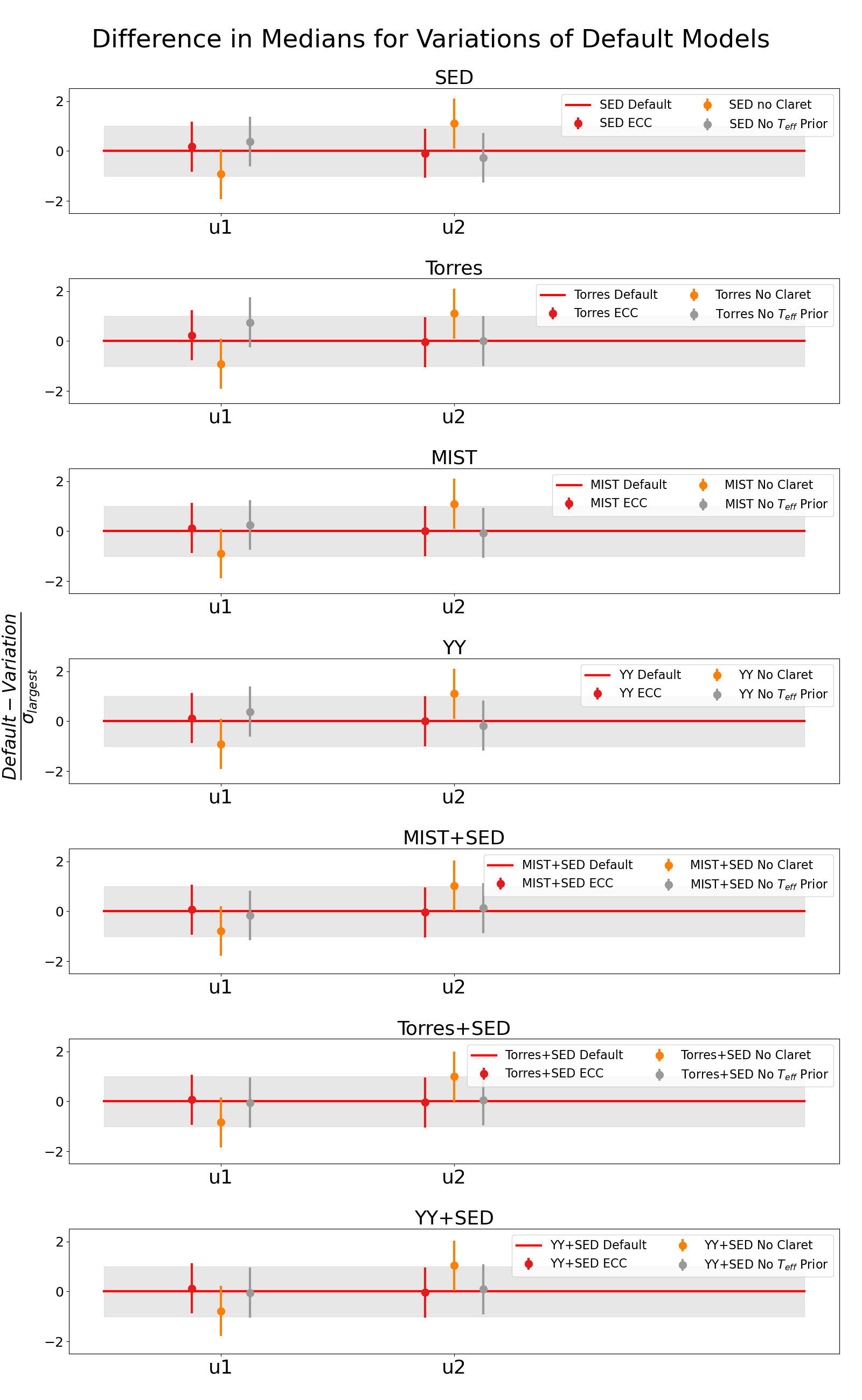}
    \caption{The color coded points represent estimates for the quadratic limb darkening coefficients, u1 and u2, in the TESS band-pass based on the stellar constraints and assumptions applied in each iteration. The LDC estimates derived from the ``default" model constraints are represented by the red line with its associated 1$\sigma$ region shown by the gray band. We take the difference of a ``variation" model and the default and divide by the largest uncertainty of all the variation models in that parameter to find the $\sigma$ difference represented here.}\label{fig:LimbDarkening}
\end{figure}

Thus, when we replace our prior on the LDCs based on the Claret Tables \citep{claret_bloemen,claret2017} with an uninformative uniform LDC prior, we find $1 \sigma$ level differences in the direct observables T, $\tau$, and $\delta$ (Fig. \ref{fig:VariationMedDiff}) relative to the fits with the Claret LDC priors.  Figure \ref{fig:LimbDarkening} shows the inferred values of the LDCs from the default fits with a LDC prior compare to those with an uninformative prior for the TESS observations showing up to $\sim 1 \sigma$ differences. We do not show the LDCs for the ground based observations as they had much larger uncertainties and thus only weakly impact the inferred parameters. 

However, we do not see any difference in the inferred values of the LDCs due to
the stellar characterization method used. This suggests that the differences in the LDCs inferred with and without the prior are because the light curve shape is not perfectly well described by the Claret LDCs. \citet{patel2022} find that model atmospheres predict values of u2 that differ by as much as $\sim$0.2 compared to those found empirically from direct fitting to precise light curves. We conclude that the choice of whether to include a LDC prior is a minor source of systematic uncertainty in the inferred parameters of transiting planet systems.

%Old figure
%May24_SEDcombo_LimbDark3.png

\subsection{Impact of Free Eccentricity}\label{sec:VariationEcc}

The value of $\rho_{*}$ inferred from the light curve and RV data depends on the eccentricity $e$ and argument of periastron of the orbit (see Eq. \ref{eqn:aoverrstar} and \ref{eqn:rhostar}). Thus, a weak constraint on $e$ leads directly to a weak constraint on $\rho_{*}$ from the data alone. Indeed, in the absence of a constraint on $e$, such as the case in the absence of RV data, single-constraint fits do not yield a unique solution for $\rm M_{*}$ and $\rm R_{*}$. Conversely, if one imposes the strict prior that $e = 0$, the constraint on $\rho_{*}$ from the light curve data becomes tighter and thus the uncertainties on the physical parameters decrease.  In the case of double-constraint models with no $e$ constraint, there is a unique solution. The SED+Gaia+\teff\ prior yields an estimate of $\rm R_{*}$, and this combined with priors on \teff\ and [Fe/H] prior and the (MIST, YY, Torres) models uniquely constrains $\rm M_{*}$ and $\rm R_{*}$. However, the system is not over-constrained in the double constraint model, as was the case we adopted $e=0$ or when there is a constraint on $e$ from RV data. It follows then that the double constraint models provide a constraint on $e$ that is independent of the RV data.

%Old Figures
%May25_ecc_paper_single_part1.png
%May25_ecc_paper_combo_part1.png

\begin{figure*}
    \centering
    \includegraphics[width=8.5cm]{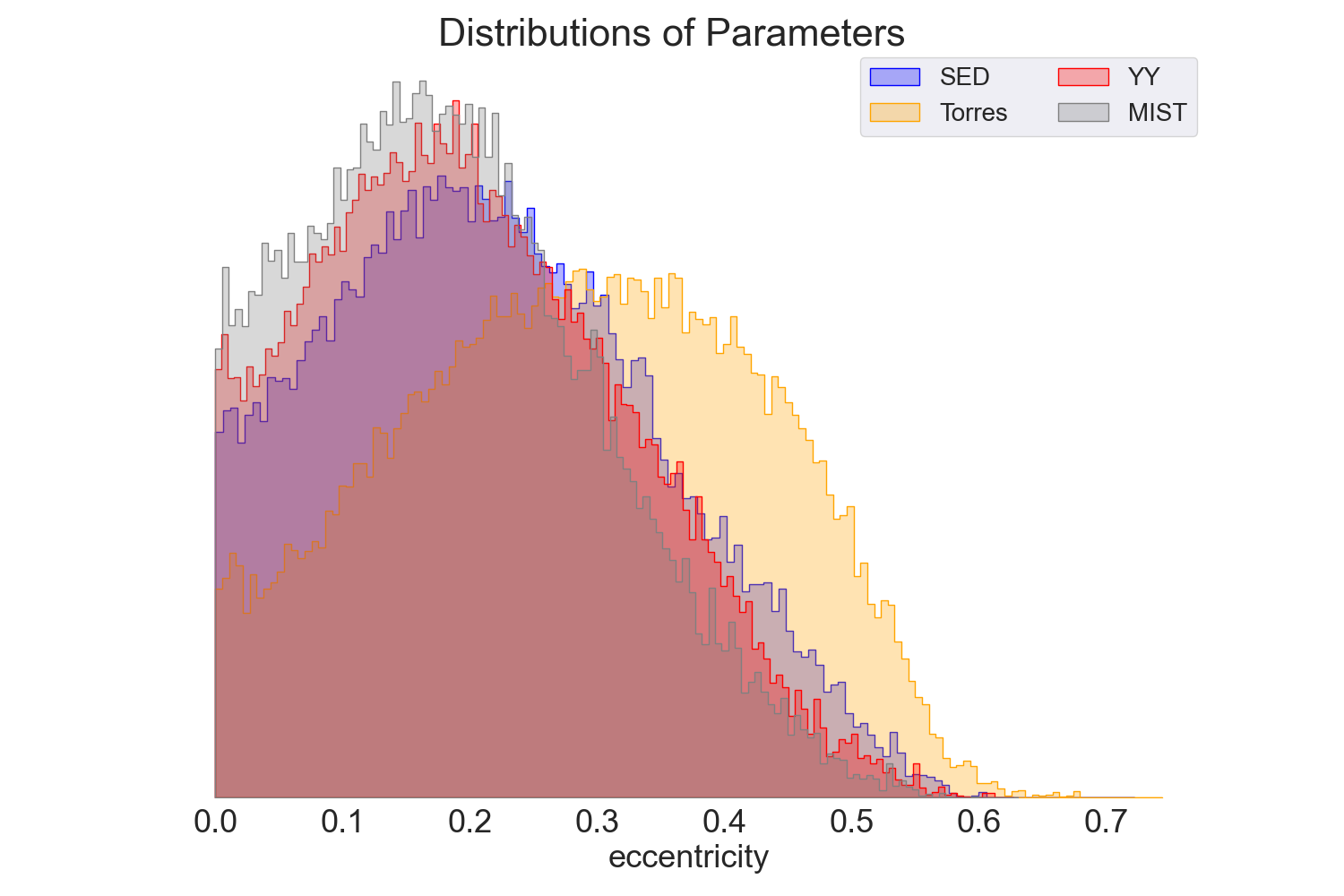}
    \includegraphics[width=8.5cm]{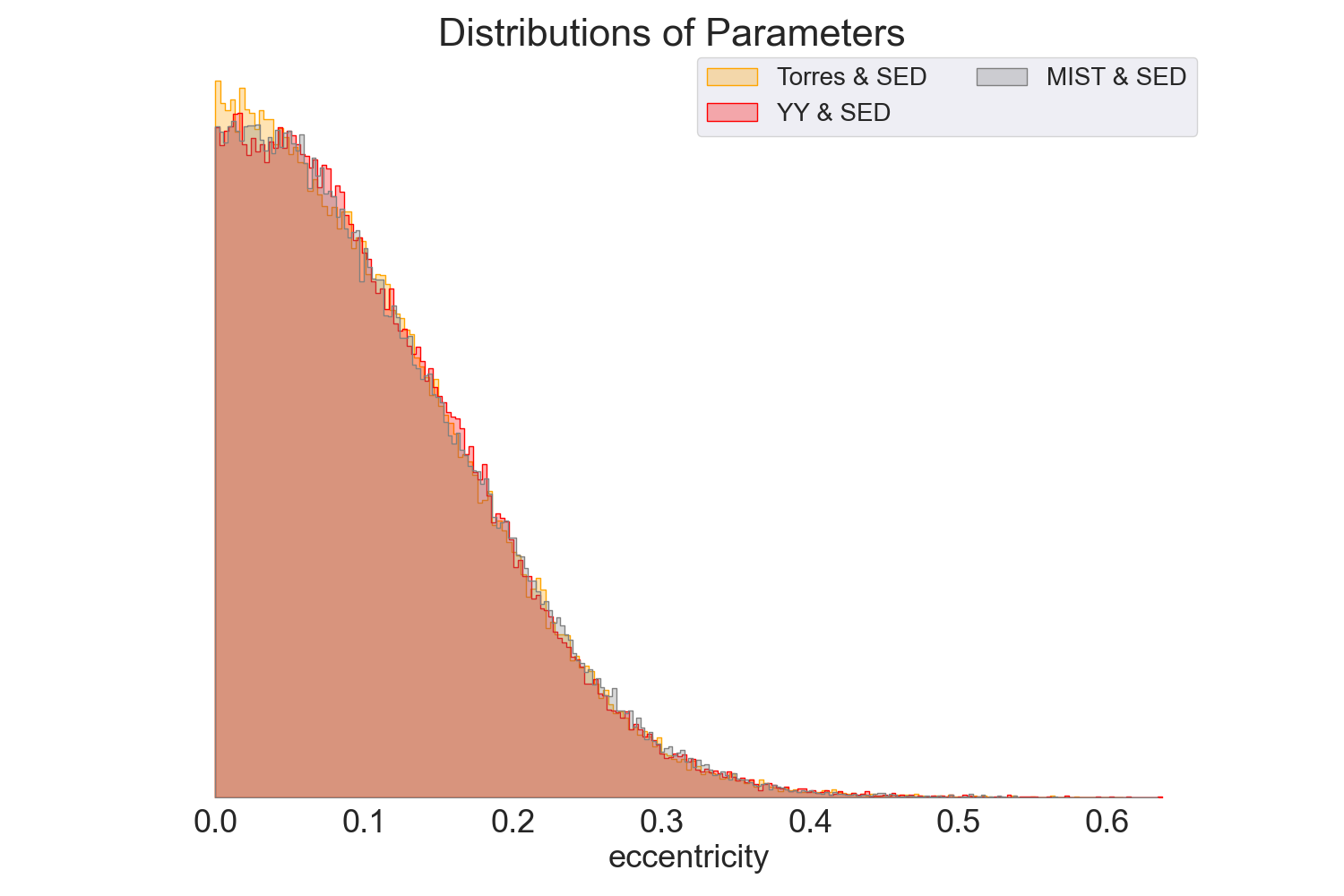}
    \caption{ The shaded histograms represent posterior distributions for the free eccentricity fit. The right panel displays the eccentricity posterior distribution for the single constraint stellar models and the left displays the combination constraint stellar models.}\label{fig:Eccentriciy}
\end{figure*}
%ecchist.py

Hot Jupiters are often assumed to have circular orbits, and a strict $e = 0$ prior is imposed. The impact of imposing $e=0$ versus using the RV data to constrain $e$ is particularly important for typical Hot Jupiters for two reasons. Then, the fractional uncertainty on $e$ is generally larger for smaller $e$ because the shape of the RV curve is weakly dependent on $e$ for $e \la 0.3$.  Second, most Hot Jupiter systems only have RV data from the discovery papers, which are sufficient to confirm the planetary nature of the companion and provide an estimate of its mass, but insufficient to provide a precise constraint on $e$.

The inferred parameters for the single constraint models assuming a uniform prior on $e$ from (0,1) are provided in Tables \ref{tab:eccCanonical} and \ref{tab:eccCanonicalCont}. The parameters for the double constraint models are presented in Tables \ref{tab:eccCanoncialCombo} and \ref{tab:eccCanoncialComboCont}. In Figure \ref{fig:Eccentriciy}, we show the posterior distributions of the eccentricity for the single and double constraint models assumptions.   In all cases, we find that the eccentricity is consistent with zero to with the 2.5 sigma threshold needed for a significant detection of eccentricity considering  the \citet{LucySweeney1971} bias. We note that the distribution of eccentricities differs from the single and double constraint models, with the latter being more consistent with zero. This difference arises because of the constraint on $e$ provided by the double constraint models. 

We find that the uncertainties in the inferred physical parameters increase when relaxing the circular prior, as expected. As expected, it has a particularly large effect on the median value and uncertainty in the stellar density. This is then reflected in the uncertainties in the other physical parameters. For example, the statistical uncertainty in $\rm R_{*}$ goes from $2.0\%$ in the “default" MIST+SED combination model to $3.6\%$ in the eccentric MIST+SED model. There is a similar increase in uncertainty in $\rm R_{p}$. Thus, imposing the constraint of a circular orbit is a source of systematic uncertainty in the precision with which the physical parameters are inferred, and adopting $e=0$ may artificially constrain the uncertainty ranges for these other inferred parameters. We therefore encourage caution in its use.

\subsection{Impact of Adopting a Prior on $\rm T_{eff}$}\label{sec:Teffprior}

Our default models impose a prior on \teff\ from the measurement by \citet{sweetcat} using high-resolution spectra.  We adopt a $2\%$ error floor as recommended by \citet{tayar2022}. For the single-constraint MIST, YY, and Torres fits, this prior on \teff\ is required to yield a complete solution to the system. Without a \teff\ prior, the only constraints on the properties of the host star are the spectroscopic [Fe/H] prior, and the stellar density $\rho_*$ inferred from the light curve and RV data. Thus the fact that the age is unconstrained means that there is a one parameter degeneracy in $\rm M_{*}$ and $\rm R_{*}$, with \teff\ being the free parameter, which becomes a proxy for age. In fact, this is not strictly true, as it ignores the constraint on \teff\ from the shape of the light curve through the LDCs. However, this constraint is relatively weak (see S \ref{sec:VariationClaret}).

\begin{figure*}
    \centering
    %Made with ParameterScatterPlots.py
    \includegraphics[width=16cm]{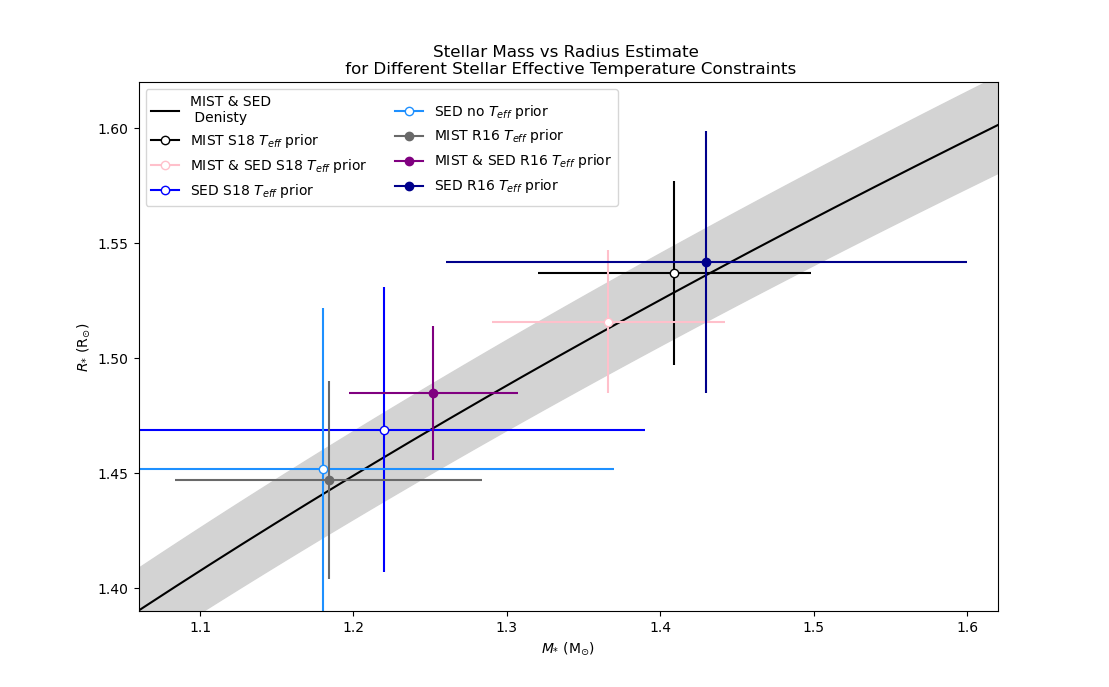}
    \caption{ Mass and radius estimates for the host star using a \teff\ prior from \citet{kelt15b} (shortened to R16) are represented by the filled points. Mass and radius estimates for the host star using a \teff\ prior from \citet{sweetcat} (shortened to S18) are represented by the open points. We see that the fit using the SED constraint and the \citet{sweetcat} \teff\ prior agrees closely with the SED fit that does not adopt a \teff\ prior. 
    }\label{fig:Teff_star_radius}
\end{figure*}

There is also a subtle difference between the single-constraint Torres fits and the single constraint YY and MIST fits. For the Torres fits with an uninformative prior on \teff, the problem is mathematically under-constrained. That is, the analytic Torres relations describe $\rm M_{*}$ as a function of $\log{g_*}$, \teff, and [Fe/H], and similarly $\rm R_{*}$ as a function of $\log{g_*}$, \teff.  With no \teff\ constraint, there are five unknowns ($\rm M_{*}$, $\rm R_{*}$, $\log{g_*}$,  \teff\ and [Fe/H],  two constraints ($\rho_*$, [Fe/H]), and two equations, and thus there is a mathematical one-dimensional degeneracy.  In the case of YY and MIST, \teff\ becomes a proxy for the age of the star, but the age of the star is not strictly unbounded. Thus the problem is only partially under-constrained. As an example of this subtlety, we find that the MIST single-constraint models with an uninformative prior on \teff\ result in smaller statistical uncertainties in the parameters than for the Torres fits.

For the single constraint SED fits, a constraint on \teff\ is needed to infer $\rm R_{*}$ from the bolometric flux and parallax. Without this constraint, there is a one-parameter degeneracy in $\rm R_{*}$ with \teff, and thus a one parameter degeneracy in $\rm M_{*}$ when combined with the constraint on $\rho_*$ from the light curve and RV data. While high-resolution spectra typically provide the most precise measurements of \teff, the SED itself also provides a constraint on \teff, albeit a weaker one. Thus from the double constraint (MIST, YY, Torres)+SED fits, it is possible to obtain a unique solution to the system with an external prior on \teff, although in this case the problem is not over-constrained, as it is when such a prior is imposed. Again, this ignores the weak constraint on \teff\ from the shape of the light curve through the LDC.

We generally find that the statistical uncertainties in the inferred system parameters are larger without the spectroscopic \teff\ prior, particularly for the single constraint fits. This would suggest that obtaining a spectroscopic measurement of \teff\ is essential to achieve the best precision in the inferred system parameters. However, it also implies that any systematic error in the value of \teff\ will propagate to the system parameters. 

In order to explore possible systematic errors imposed by an inaccurate value for our spectroscopic \teff\ prior, we compare the SED default model fit to that with an uninformative prior on \teff\ in Figure \ref{fig:Teff_star_radius}. We see broad agreement between our “default" model with the SED constraint and the variation with an uninformative uniform prior on the stellar effective temperature at the level of $0.24 \sigma$. This indicates agreement between the SED derived \teff\ estimate and the spectroscopic \teff\ estimate from \citet{sweetcat}. On the other hand, we find a difference of $\sim 1.3 \sigma$ in the \teff\ estimate from the SED fits without a spectroscopic prior relative to the original \citet{kelt15b} \teff\ estimate, suggesting that the \citet{kelt15b} \teff\ estimate may be inaccurate and/or have an underestimated uncertainty. Similarly, we see a $2 \sigma$ difference in the stellar radii estimate between the MIST and SED fits which use the \citep{kelt15b} \teff\ prior. This gap narrows to $\sim 1\sigma$ between the MIST and SED fits when we adopt the \citet{sweetcat} \teff\ prior.

We find several interesting interactions between the estimates of age, luminosity, stellar effective temperature, and extinction when comparing the fits using spectroscopic priors from \citet{sweetcat} (S18) or \citet{kelt15b} (R16).

We see extinction play a larger role when comparing SED R17 and SED S18 which produces a $1 \sigma$ difference in $\rm R_{*}$. The spectroscopic \teff\ prior for SED R17 is $\sim 400$K cooler and \exofastv\ fits an extinction close to zero ($0.1 \pm 0.1$). Thus \exofastv estimates a luminosity of $\rm L_{*} \sim 2.87 L_{\odot}$. The spectroscopic \teff\ prior for SED S18 is hotter and \exofastv\ estimates extinction of $0.3 \pm 0.1$. Thus \exofastv estimates that SED S18 would have a larger luminosity ($\rm L_{*} \sim 3.36 L_{\odot}$). So SED R16 is $94\%$ of the S18 \teff, less extincted, and $86\%$ of the S18 luminosity. Given that $\rm R_{*} \propto \frac{\sqrt{L_{*}}}{\rm T_{eff}^2} $, the decrease in temperature compared to S18 dominates and the leaves the SED R16 $\rm R_{*}$ estimate $\sim 1.1\sigma$ (or $\sim 4\%$) larger than the SED S18 estimate.

When we compare the MIST R16 and MIST S18 we find a $\sim 2 \sigma$ difference in $\rm R_{*}$ due to the 400K or $3 \sigma$ difference in the
temperature priors and $\sim2 \sigma$ difference in the [Fe/H] priors. These different sets of priors leads to the MIST models estimating different evolutionary positions for the host star. MIST R16 finds an $\rm L_{*}$ estimate $70\%$ that of the MIST S18 fit and due to the difference in priors a \teff estimate $94\%$ of the MIST S18 estimate.  The decrease in luminosity dominates over the decrease in temperature, and as a result the MIST S18 inferred $\rm R_{*}$ is $\sim 2 \sigma$ or $6\%$ larger than the MIST R16 inferred $\rm R_{*}$.

Generally, the MIST and MIST+SED fits have a smaller gap than the SED and MIST+SED fits for a single set of priors. The smaller uncertainties in MIST dominate, so the MIST+SED is weighed more closely to the MIST estimates. Thus, the disparate nature of Figure \ref{fig:Teff_star_radius} is due to the degeneracies between stellar effective temperature, luminosity, and extinction.

%%%%%%%%%%%%%%%%%%%%%%%%%%%%%%%%%%%%%%%%%%%%%%%%%%%%%%%%%%%%%%%%%%
\section{Discussion} \label{sec:discussion}

In many ways, KELT-15b \citep{kelt15b} is a typical hot Jupiter. KELT-15b has a bright F-type host star and a planetary radius comparable to Jupiter. There are many photometric transit observations, particularly from \tess. However, the radial velocity observations are limited to those presented in the discovery paper, used to confirm the planetary nature of the companion and provide only a rough estimate of the planet mass. The characteristics of this planet and its suite of observations make it a good proxy for most known hot Jupiters. Observations from \tess\ will continue to bolster this population by finding even more large planets around bright stars. In general, these planets will also have relatively few ground-based follow up radial velocity observations. The re-analysis of KELT-15b provides an excellent test-bed with which to explore the systematic biases present in the modeling of this population of transiting planets.

We find that broad consistency in the host star and planetary parameters inferred using different methods of breaking the stellar mass-radius degeneracy can be significant. For example the largest differences we find between the inferred stellar radii using different methods are at the $\sim 3\%$ level or roughly the magnitude of the largest statistical uncertainty. This translates directly into a similar $\sim 1\sigma$ difference in the planetary radius. Given that we are entering an era of precision exoplanetology, where precise {\it and} accurate physical parameters of transiting planets are needed for planning and interpreting follow-up observations, broad agreement between differing methodologies used is the community is required to draw meaningful conclusions from population studies of planets. Generally, we encourage specific reporting of any explicit or implicit prior assumptions in the fitting of transiting exoplanet systems and of the methodologies used to break the mass-radius degeneracy in the host stars in order to precisely characterize the properties of these systems. In an era where the analysis of planetary interiors requires radii and densities accurate to a few \%, a $3\%$ source of systematic uncertainty is approaching the level of a notable consideration.

Although we investigate the impacts of several different methods of breaking the mass-radius degeneracy in the host star and variations in a select set of our modelling assumptions, our study does not explore all possible sources of systematic error. For example, even within a given method of breaking the mass-radius degeneracy in the primary star, there can be significant variations in the inferred parameters. For example, \citet{ARIADNE} uses Bayesian Model Averaging in the python package ARIADNE to improve the SED fitting technique through incorporating information from 6 different stellar atmosphere models. Within these stellar atmospheres used for SED fitting, the authors find individual stellar atmosphere models can vary by as much as 32\% in radius and 12\% in \teff. They find their model averaging technique has a fractional difference of 0.001 $\pm$ 0.070 and has a mean precision of 2.1\%. This shows that there can be significant improvements made to current methods of SED fitting techniques, which is just one method of breaking the mass-radius degeneracy.

The photometry included in the SED fits has also dramatically increased in volume in recent years. \citet{yu2022} combined the data-sets from APOGEE \citep{apogee2022}, GALAH \citep{galah2021}, and RAVE \citep{rave2020} to revise the radii for 1.5 million stars using the SED characterization method on these photometric data-sets. They were able to use 32 photometric bands and the MARCS \citep{marcs} and BOSZ \citep{bosz} model stellar atmospheres to fit the SEDs. With this detailed set of photometry for a large sample of targets, the researchers obtained radii that were consistent to 3\% with CHARA \citep{chara} interferometric radii. Overall their catalogue has a median precision of 3.6\% in stellar radius. Observations from \textit{Gaia} DR3 \citep{Gaiadr3} and beyond will further enhance the widely available SED photometry by providing spectroscopic atmospheric parameters for 5.5 million stars.

%references to the Tayar paper
Further studies have investigated the systematic uncertainties introduced by inhomogenous methods of breaking the stellar mass-radius degeneracy. Our results of a difference of $\sim 3\%$ or $\sim 1\sigma$ in $R_{*}$ between several degeneracy breaking methods are closely in line with the results of \citet{tayar2022}. These authors find a floor of approximately 4\% in $R_{*}$ using theoretical estimates of solar type stars by comparing 4 different stellar evolution codes. The authors simulate populations of stars at fixed masses and metallicities. That work examines the Yale Rotating Evolution Code Models (YREC) \citep{yrec}, the MIST isochrones \citep{mist}, the Dartmouth Stellar Evolution Program (DSEP) \citep{dsep}, and the Garching Stellar Evolution Code (GARSTEC) \citep{garstec1, garstec2}. The only overlapping model with our analysis  and in \citet{tayar2022} is the use of the MIST isochrones \citep{mist}, showing that we find similar results for a different suite of stellar modeling methods. The authors consider the ``maximal difference" between all four models as the estimate of systematic error. \citet{tayar2022} find offsets in stellar mass estimate of $\sim 5\%$ on the main sequence and sub-giant branches and as large as $\sim 10\%$ near the base of the red-giant branch. Here the metallicities and \teff\ values are held constant. They find differences of $\sim 7\%$ where the approach to convective overshoot is relevant near masses $\sim 1.1 M_\odot$, showing the impact of stellar physics assumptions. \citet{tayar2022} consider the exoplanet hosting systems $\pi$ Mensae \citep{pimen2018} and TOI-197 \citep{toi197}. In both cases they find the systematic uncertainty in the host star mass to be at the same level as the statistical uncertainty.

Recent work in \citet{eastman2023} shows that precise estimates of stellar density can be used to beat the systematic error floors in radius and effective temperature \citep{tayar2022}. They assume a 5\% systematic floor on estimates of $M_{*}$ (in line with the 5\% uncertainty achieved in this work in the combination constraint models) and a density measurement of 1\% (we achieve $\sim 2\%$ in the combination constraint models) providing an estimate of precision in $R_{*}$ of 1.7\%, agreeing with the precision achieved in this work for the combination constraint models. \citet{eastman2023} also show the power of this method to estimate \teff\ from precisely constrained stellar densities. \citet{eastman2023} then uses these assumptions to estimate systematic errors in the WASP-4b system, whereas we model the KELT-15b systems under several different assumptions to characterize its systematic uncertainties.  \citet{eastman2023} also shows \teff\ could be constrained to roughly 1.5\% assuming a 5\% uncertainty in $M_{*}$ and 5\% in $\rm \rho_{*}$.

While the results of \citet{tayar2022} and \citet{ARIADNE} explore the inference of different stellar properties inferred using different stellar evolution and atmosphere models, we are able to see the full impact of these differences through modeling an entire system several times with different constraints. By globally modeling the system, we are capable of incorporating additional constraints on $\rho_{*}$ provided by the transit observations themselves.

This analysis and similar results in the literature support detailed reporting of stellar characterization methods and prior assumptions when considering the total error budget of transiting exoplanet systems \citep{tayar2022,eastman2023}. These systematic differences between the parameters inferred by different methods of breaking the mass-radius degeneracy and different prior assumptions make comparisons of exoplanets that have been modeled using inhomogeneous methods nontrivial. When drawing comparisons between exoplanetary systems, it is important to account for the full error budget of these systems by including these systematic uncertainties derived from the selection of the stellar characterization technique and the selection of model priors. Of course, this is complicated by the fact that these systematic errors are typically unquantified, something which we have aimed to rectify here.

\section{Best Practices and Conclusions}\label{sec:conclusions}

Using the KELT-15 transiting Hot Jupiter systems \citep{kelt15b} as a test bed, we quantify, for the first time, the systematic differences in the inferred physical properties of the host star and planet due to the choice of several prior assumptions. In particular, we consider four different methods of breaking the well-known mass-radius degeneracy in transiting planet systems: a SED in combination with a Gaia parallax, MIST isochrones \citep{mist}, YY isochrones \citep{YY2001}, and the empirical Torres relations \citep{Torres_2009}. We also consider combination constraints including the SED+YY, SED+MIST, and SED+Torres. Finally, for each of these seven methods, we consider the impact of three other priors compared to a ''default", namely the limb-darkening coefficient, the stellar effective temperature, and the eccentricity.  In total, we explore the impact of 28 different sets of assumptions for the priors on the inferred parameters of KELT-15b. Our primary conclusions are:

\begin{itemize}

\item For systems with radial velocity datasets with relatively poor statistical precision, the choice of the prior used to sample the velocity semi-amplitude (e.g., $\log{K}$ vs. $K$), or the planet mass ($\log{M_{p}}$ versus $M_{p}$), can result in posteriors with non-negligible support for very low $K$ or $M_p$, which in turn impacts other inferred parameters.- \S \ref{sec:SanityCheck}\\

\item The inclusion of \tess\ data improves our precision on planet period by 50\% compared to \citet{kelt15b} - \S \ref{sec:results}\\

\item Our single constraint models display agreement in stellar and planetary radius at the a 4.4\% or a $1.1\sigma$ level between the smallest estimate found using the SED constraint and the largest estimate found using the MIST constraint.- \S \ref{sec:DefaultResults}\\

\item We find strong agreement (better than 0.2$\sigma$) in stellar and planetary radii among our combination constraints (MIST+SED, YY+SED, Torres+SED) - \S \ref{sec:ComboResults} \\

\item Simultaneously fitting the host star and planet allow for stellar properties like $\rm \rho_{*}$ to be constrained by both direct observable parameters and independent estimates from \teff\ and the mass-radius degeneracy breaking method used - \S \ref{sec:VariationClaret}\\

\item We find that the decision to employ or not employ a prior on our limb darkening coefficients based on the Claret tables \citep{claret2017} provides a modest but noticeable effect on the ``directly" observable parameters at the $1 \sigma$ level - \S \ref{sec:VariationClaret}\\

\item Even in systems with eccentricities that are consistent with zero, adopting a prior that $e=0$ may significantly- and perhaps artificially - reduce the statistical uncertainties on many parameters. - \S \ref{sec:VariationEcc}\\

\item The estimates of stellar properties depended significantly on the adoption of a spectroscopic \teff\ prior.  We find a $\sim 2 \sigma$ difference in the $R_*$ estimated by the MIST models when we adopt differing literature spectroscopic \teff\ estimates. We suggest comparison to a fit using the SED and no temperature prior when deciding to use a spectroscopic \teff\ prior. - \S \ref{sec:Teffprior}\\

\end{itemize}

Based on the results of our analysis, we offer for the community's consideration several suggestions for ``best practices" when analyzing transiting planet systems. We find that the inclusion or exclusion of a prior on limb darkening coefficients based on limb darkening models has notable impacts on direct observables like the measured transit depth and duration. Generally, we find most of the models are very sensitive to whether or not a prior on \teff\ is adopted. We strongly recommend careful consideration of the constraints on \teff\ adopted in studies of transiting systems. The MIST and SED combination is the most robust against this change. We find that the use of the Claret tables to fit limb darkening coefficients compared to fitting these parameters freely has the largest impact on the determination of the transit duration, and other directly observed parameters due to tensions in fitting $\rho_{*}$. We find that, even when including the SED, whose shape provides an independent constraint on \teff, applying the spectroscopic constraint on \teff\ results in smaller statistical uncertainties on $M_*$, $R_*$, and $R_p$. However, systematics in the spectroscopically-measured \teff\ should be considered \citep{tayar2022} and may imply that the prior on the spectroscopically-derived \teff\ should not be included \citep{eastman2023}.

When SED information is available, we recommend combining it with another method of breaking the host star mass and radius degeneracy. SED alone produces one of the least precise estimates for the mass and radius of the host star. When the SED constraint is combined with the YY constraint,  we see a precision increase by a fractional difference of $\sim 33$\% in the stellar radius, while still retaining the benefits of this empirical constraint. Additionally we see strong agreement between the models that incorporate SED and parallax information, providing consistent estimates of stellar radii. Even when eccentricities are consistent with zero, we see significant increases in uncertainty for several derived parameters like $R_{*}$ and $R_{p}$. By applying the constraint of a circular orbit as a matter of course, the uncertainties in these derived parameters could be artificially constrained. This is an important consideration when attempting an accurate accounting of the error budget of a planetary system.

This work also highlights the importance of globally modeling both the host star and transiting planet simultaneously. We have found several instances where changes in the information being applied to the host star such as providing a prior on \teff, the method used to fit limb darkening coefficients, or combining indirect stellar characterisation methods with the SED, produce significant differences in the directly observable and the inferred properties of the transiting planet. Many of these effects would not have been considered if we had sequentially and independently modelled the host star then the transiting planet. For example, fitting the host star independently from the planet completely elides the role the host star's properties play in fitting the ``directly" observable parameters. We would not consider the ability of the light curve direct observables to assist in constraining $\rho_{*}$ rather than solely relying on the derived estimate from the effective temperature. This important independent constraint would then be lost. As discussed in \S \ref{sec:StellarModels}, the tension between the $\rho_{*}$ estimated by \teff\ and the $\rho_{*}$ estimated by the light curve are significant factors in fitting the direct observables. These observables in turn are then used to estimated the derived quantities like $R_{*}$ and $R_{p}$. Thus we incorporate far more information to accurately characterise the host star and transiting planet when we model these systems simultaneously and globally.

In the future, a consistently agreed-upon method for breaking the mass-radius degeneracy is essential to the comparative study of exoplanets. This would reduce the role of systematic errors when making detailed comparisons between exoplanets. If a consistent model is not used, detailing which host star mass-radius degeneracy breaking method is used is necessary to be able to account for the systematic differences outlined in this work. In the future, we hope to apply this analysis to a set of benchmark exoplanet hosting systems, i.e., bright host stars with observations from \textit{Gaia} \citep{gaia} or SPHEREX \citep{spherex}, interferometrically measured radii, asteroseismologically-inferred masses and radii, and those with "flicker"-based estimates of $\log{g_*}$. This would allow us to compare these models and prior assumptions to not only each other but to an approximation of the ground truth for the stellar and planetary properties. This would provide a pathway to defining a method of analyzing transiting planets that is most consistent with ground truth.

\section*{Acknowledgements}

A.D. and B.S.G. were supported by the Thomas Jefferson Chair for Space Exploration endowment from the Ohio State University. A.D. would also like to thank Romy Rodr{\' i}guez Mart{\' i}nez and David V. Martin for conversations and technical insight.

%http://astrofrog.github.io/acknowledgment-generator/
Some/all of the data presented in this paper were obtained from the Mikulski Archive for Space Telescopes (MAST). STScI is operated by the Association of Universities for Research in Astronomy, Inc., under NASA contract NAS5-26555. Support for MAST for non-HST data is provided by the NASA Office of Space Science via grant NNX13AC07G and by other grants and contracts. This research made use of NumPy \citep{harris2020array}.

\section*{Data Availability}
The specific observations analyzed can be accessed via \href{https://doi.org/10.17909/sh0j-ym03}{10.17909/sh0j-ym03}.  Other data is available upon request.  \exofastv \citep{exofastv2} can be publicly accessed via \href{https://github.com/jdeast/EXOFASTv2}{https://github.com/jdeast/EXOFASTv2}

\bibliographystyle{mnras}
\bibliography{papers}

\section*{Appendix A}\label{app:A}

To understand the nature of the mass-radius degeneracy in transiting planet systems, we first consider the ``directly observable'' parameters from relative photometry and radial velocities. Directly observable parameters \footnote{We consider direct observables to be parameters that can essentially be determined from a given data set alone, without combining additional information (data or models) from external sources.  In many cases these observables can be 'read off' of the data, i.e., estimated from visual inspection, and are generally weakly dependent on the choice of priors. .  For example, the period of a transiting planet can be more-or-less read off from a light curve, whereas the semi-major axis cannot. } are those that can be determined solely from photometric and RV observations \citep{Seager_2003,carter2008,Winn2010}. Such parameters include the planet period $P$, transit depth $\delta$, the transit duration defined by full width at half the maximum depth  $T$, the ingress/egress duration $\tau$, and the eccentricity $e$, argument of periastron $\omega$ and semi-amplitude $K_*$ of the RV observations. From these direct observables, we can determine the semi-major axis of the planet orbit scaled to the radius of the star ($a/R_*$), as well as the surface gravity of the planet $g_P$.

The RV semi-amplitude $K_\star$ can constrain the planet mass $M_p$, given an estimate of $M_\star$ via 
\begin{equation}
    K =\left(\frac{2\pi G}{P}\right)^{1/3} \frac{M_p\sin{i}}{(M_{\star}+M_p)^{2/3}} \frac{1}{\sqrt{1-e^2}},
    \label{eqn:K}
\end{equation}
where the precise shape of the RV curve can be used to measure the eccentricity $e$. The shape of the light curve constrains the inclination of the orbit $i$.  Thus by measuring $K_\star$, we establish a relationship between $M_{*}$ and $M_{p}$ which depends only on the observable parameters $K$, $e$, $i$ and $P$.

Similarly, the transit depth provides a constraint on the planet radius $R_P$ when $R_*$ is known through
\begin{equation}
    \delta = \left(\frac{R_P}{R_\star}\right)^2.
\end{equation}
This depth can be measured more-or-less directly from the relative photometry of the host star during transits,ignoring limb darkening, the impact of which we explore in S \ref{sec:StellarModels}. Thus $\delta$ serves as a direct observable that establishes a relationship between $R_{*}$ and $R_p$. 

Direct observables also constrain the ratio of the semi-major axis a and $R_{*}$,
\begin{equation}
    \frac{a}{R_*} =  \frac{\delta^{1/4}}{\pi} \frac{P}{\sqrt{ T \tau}} \left(\frac{\sqrt{1-e^2}}{1+e\sin\omega}\right).
\label{eqn:aoverrstar}
\end{equation}
Using this dimensionless factor, one can derive a relationship for the density of the host star $\rho_{*}$ and the surface gravity of the planet $g_{p}$ solely in terms of direct observables. The density of the star is given in \citet{Seager_2003} by 
\begin{equation}
\rho_\star = \frac{3\pi}{GP^2}\left(\frac{a}{R_\star}\right)^3
-k^3\rho_p \simeq \frac{3\pi}{GP^2}\left(\frac{a}{R_\star}\right)^3,
\label{eqn:rhostar}
\end{equation}
\noindent where $k\equiv R_p/R_*$. For transiting planets, $k\ll1$ and $\rho_p \sim \rho_*$, justifying the second approximation. The surface gravity of the planet is given in \cite{southworth},
\begin{equation}
    g_{p} = \frac{2K_{\star}P}{\pi T\tau} \delta^{-1/2}.
\label{eqn:gpobservables}
\end{equation}

However, it is not possible to estimate absolute values for $R_*$ and $M_*$ through the relative photometric measurements of the transit or the RV observations alone. 
These relationships leave a one-parameter degeneracy between $R_{*}$ and $M_{*}$. Such a degeneracy implies that the mass and radius of the exoplanet can not be determined with only photometric and RV observations. Additional information must be introduced to break this degeneracy and fully characterise both the host star and the exoplanet. The specific method used to break this degeneracy can then introduce unquantified systematic uncertainties.

\section*{Appendix B}\label{app:B}

%%%%%%%%%%%%%%%%%%%%%%%%%%%%%%%%%%%%%%%%%%%%%%%%%%%%%%%%%%%%%%%%%%%%%%%%%%%%%%%%%%%%%%%%%%%%%%%%%%%%%%%%%%%%%%%%%
\setlength{\extrarowheight}{3pt}
\begin{table*}
	\centering
	\caption{Results for circular single constraint models. The maximum difference between the models divided by the largest uncertainty is described by $\sigma$.}
	\label{tab:CanonicalTable}
\begin{tabular}{lcccccl}
	\toprule	& MIST & YY & Torres & SED & $\sigma$ & Units	 \\
	\toprule
	$M_*$& $ 1.409^{+0.086}_{-0.089} $& $ 1.391^{+0.057}_{-0.056} $& $ 1.382^{+0.074}_{-0.071} $& $ 1.22^{+0.17}_{-0.15} $& $1.18$& $(\msun)$	\\
	$R_*$& $ 1.537^{+0.04}_{-0.038} $& $ 1.53^{+0.033}_{-0.028} $& $ 1.528^{+0.037}_{-0.033} $& $ 1.469^{+0.062}_{-0.059} $& $1.12$& $(\rsun)$	\\
	$L_*$& $ 3.65^{+0.41}_{-0.38} $& $ 3.66^{+0.39}_{-0.36} $& $ 3.64^{+0.39}_{-0.36} $& $ 3.36^{+0.34}_{-0.3} $& $0.76$& $(\lsun)$	\\
	$\rho_*$& $ 0.55^{+0.016}_{-0.026} $& $ 0.55^{+0.016}_{-0.026} $& $ 0.549^{+0.017}_{-0.027} $& $ 0.549^{+0.017}_{-0.026} $& $0.05$& (cgs)	\\
	$\log{g}$& $ 4.214^{+0.012}_{-0.015} $& $ 4.2124^{+0.01}_{-0.014} $& $ 4.21^{+0.012}_{-0.015} $& $ 4.193^{+0.023}_{-0.026} $& $0.86$& (cgs)	\\
	$T_{\rm eff}$& $ 6430.0^{+120.0}_{-120.0} $& $ 6450.0^{+120.0}_{-120.0} $& $ 6450.0^{+120.0}_{-120.0} $& $ 6450.0^{+120.0}_{-120.0} $& $0.17$& (K)	\\
	$[{\rm Fe/H}]$& $ 0.233^{+0.079}_{-0.08} $& $ 0.236^{+0.08}_{-0.08} $& $ 0.236^{+0.079}_{-0.079} $& $ 0.225^{+0.078}_{-0.078} $& $0.14$& (dex)	\\
\hline
	\end{tabular}
{\raggedright  $^{1}$ Optimal time of conjunction minimizes the covariance between $T_C$ and Period \par}
\end{table*}

\begin{table*}
	\centering
	\caption{Continuation of results for the circular single constraint models. The maximum difference between the models divided by the largest uncertainty is described by $\sigma$.}
	\label{tab:CanonicalTableCont}
	\begin{tabular}{lcccccl}
	\toprule	& MIST & YY & Torres & SED & $\sigma$ & Units	 \\
	\toprule
	$\rho_P$& $ 0.409^{+0.094}_{-0.097} $& $ 0.411^{+0.093}_{-0.097} $& $ 0.408^{+0.093}_{-0.096} $& $ 0.426^{+0.098}_{-0.1} $& $0.18$& Density (cgs)	\\
	$logg_P$& $ 3.068^{+0.087}_{-0.12} $& $ 3.068^{+0.087}_{-0.12} $& $ 3.065^{+0.087}_{-0.12} $& $ 3.066^{+0.088}_{-0.12} $& $0.03$& Surface gravity	\\
	$\Theta$& $ 0.047^{+0.01}_{-0.011} $& $ 0.047^{+0.01}_{-0.011} $& $ 0.047^{+0.01}_{-0.011} $& $ 0.049^{+0.011}_{-0.011} $& $0.18$& Safronov Number	\\
	$\fave$& $ 2.07^{+0.17}_{-0.16} $& $ 2.1^{+0.17}_{-0.16} $& $ 2.1^{+0.18}_{-0.16} $& $ 2.1^{+0.17}_{-0.15} $& $0.18$& \specialcell{Incident Flux \\ (\fluxcgs)}	\\
	$T_P$& $ 2458534.08841^{+0.00013}_{-0.00014} $& $ 2458534.08841^{+0.00014}_{-0.00014} $& $ 2458534.08841^{+0.00013}_{-0.00014} $& $ 2458534.08842^{+0.00014}_{-0.00014} $& $0.07$& \specialcell{Time of Periastron \\ (\bjdtdb)}	\\
	$T_S$& $ 2458535.75314^{+0.00013}_{-0.00014} $& $ 2458535.75314^{+0.00014}_{-0.00014} $& $ 2458535.75314^{+0.00013}_{-0.00014} $& $ 2458535.75314^{+0.00013}_{-0.00014} $& $0$& Time of eclipse (\bjdtdb)	\\
	$T_A$& $ 2458536.5855^{+0.00013}_{-0.00013} $& $ 2458536.58551^{+0.00013}_{-0.00014} $& $ 2458536.58551^{+0.00013}_{-0.00014} $& $ 2458536.58552^{+0.00014}_{-0.00014} $& $0.14$& \specialcell{Time of Ascending \\ Node (\bjdtdb)}	\\
\hline
	\end{tabular}
{\raggedright  $^{2}$ the velocity of the planet at the time of transit with the estimated eccentricity, Ve, divided by the velocity the planet would have if its orbit were circular, Vc \par}
\end{table*}

%%%%%%%%%%%%%%%%%%%%%%%%%%%%%%%%%%%%%%%%%%%%%%%%%%%%%%%%%%%%%%%%%%%%%%%%%%%%%%%%%%%%%%%%%%%%%%%%%%%%%%%%%%%%%%%%%

\begin{table*}
	\centering
	\caption{Results for eccentric single constraint models. The maximum difference between the models divided by the largest uncertainty is described by $\sigma$.}
	\label{tab:eccCanonical}
	\begin{tabular}{lcccccl}
	\toprule	& MIST Eccentric & YY Eccentric & Torres Eccentric & SED Eccentric & $\sigma$ & Units	 \\
	\toprule
	$M_*$& $ 1.47^{+0.13}_{-0.12} $& $ 1.449^{+0.12}_{-0.093} $& $ 1.49^{+0.13}_{-0.12} $& $ 0.9^{+0.51}_{-0.39} $& $1.31$& $(\msun)$	\\
	$R_*$& $ 1.68^{+0.32}_{-0.2} $& $ 1.7^{+0.38}_{-0.22} $& $ 1.96^{+0.44}_{-0.37} $& $ 1.471^{+0.062}_{-0.059} $& $1.21$& $(\rsun)$	\\
	$L_*$& $ 4.4^{+1.9}_{-1.1} $& $ 4.5^{+2.2}_{-1.2} $& $ 6.0^{+3.0}_{-2.1} $& $ 3.36^{+0.34}_{-0.3} $& $1.04$& $(\lsun)$	\\
	$\rho_*$& $ 0.43^{+0.16}_{-0.16} $& $ 0.41^{+0.18}_{-0.17} $& $ 0.28^{+0.22}_{-0.12} $& $ 0.4^{+0.2}_{-0.17} $& $0.81$& (cgs)	\\
	$\log{g}$& $ 4.152^{+0.082}_{-0.13} $& $ 4.135^{+0.1}_{-0.14} $& $ 4.03^{+0.16}_{-0.14} $& $ 4.06^{+0.18}_{-0.25} $& $0.57$& (cgs)	\\
	$T_{\rm eff}$& $ 6430.0^{+120.0}_{-120.0} $& $ 6450.0^{+120.0}_{-120.0} $& $ 6450.0^{+120.0}_{-120.0} $& $ 6440.0^{+120.0}_{-110.0} $& $0.17$& (K)	\\
	$[{\rm Fe/H}]$& $ 0.233^{+0.076}_{-0.079} $& $ 0.235^{+0.081}_{-0.082} $& $ 0.236^{+0.082}_{-0.082} $& $ 0.224^{+0.077}_{-0.076} $& $0.15$& (dex)	\\
\hline
	\end{tabular}

{\raggedright  $^{1}$ Optimal time of conjunction minimizes the covariance between $T_C$ and Period \par}
\end{table*}

\begin{table*}
	\centering
	\caption{Continuation of results for eccentric single constraint models. The maximum difference between the models divided by the largest uncertainty is described by $\sigma$.}
	\label{tab:eccCanonicalCont}
	\begin{tabular}{lcccccl}
	\toprule	& MIST Eccentric & YY Eccentric & Torres Eccentric & SED Eccentric & $\sigma$ & Units	 \\
	\toprule
	$\rho_P$& $ 0.32^{+0.16}_{-0.13} $& $ 0.31^{+0.17}_{-0.13} $& $ 0.22^{+0.17}_{-0.1} $& $ 0.35^{+0.14}_{-0.12} $& $0.87$& Density (cgs)	\\
	$logg_P$& $ 3.0^{+0.13}_{-0.16} $& $ 3.0^{+0.14}_{-0.17} $& $ 2.9^{+0.17}_{-0.19} $& $ 2.99^{+0.15}_{-0.18} $& $0.56$& Surface gravity	\\
	$\Theta$& $ 0.043^{+0.012}_{-0.011} $& $ 0.044^{+0.012}_{-0.011} $& $ 0.039^{+0.012}_{-0.011} $& $ 0.051^{+0.012}_{-0.012} $& $1$& Safronov Number	\\
	$\fave$& $ 2.37^{+0.7}_{-0.48} $& $ 2.44^{+0.8}_{-0.54} $& $ 3.01^{+0.93}_{-0.83} $& $ 2.5^{+0.88}_{-0.6} $& $0.73$& \specialcell{Incident Flux \\ (\fluxcgs)}	\\
	$T_P$& $ 2458531.09^{+0.5}_{-0.25} $& $ 2458531.06^{+0.54}_{-0.22} $& $ 2458530.93^{+0.32}_{-0.12} $& $ 2458531.05^{+0.58}_{-0.21} $& $0.41$& \specialcell{Time of Periastron \\ (\bjdtdb)}	\\
	$T_S$& $ 2458535.5^{+0.22}_{-0.23} $& $ 2458535.48^{+0.23}_{-0.23} $& $ 2458535.42^{+0.24}_{-0.23} $& $ 2458535.48^{+0.23}_{-0.23} $& $0.34$& Time of eclipse (\bjdtdb)	\\
	$T_A$& $ 2458533.25^{+0.15}_{-0.17} $& $ 2458533.25^{+0.17}_{-0.18} $& $ 2458533.36^{+0.21}_{-0.19} $& $ 2458533.26^{+0.2}_{-0.19} $& $0.55$& \specialcell{Time of Ascending \\ Node (\bjdtdb)}	\\
\hline
	\end{tabular}

{\raggedright  $^{2}$ the velocity of the planet at the time of transit with the estimated eccentricity, Ve, divided by the velocity the planet would have if its orbit were circular, Vc \par}
\end{table*}

%%%%%%%%%%%%%%%%%%%%%%%%%%%%%%%%%%%%%%%%%%%%%%%%%%%%%%%%%%%%%%%%%%%%%%%%%%%%%%%%%%%%%%%%%%%%%%

\begin{table*}
	\centering
	\caption{Results for single constraint models without a spectroscopic \teff\ prior. The maximum difference between the models divided by the largest uncertainty is described by $\sigma$.}
	\label{tab:noTeffCanonical}
	\begin{tabular}{lcccccl}
	\toprule	& MIST No $\rm T_{eff}$ prior & YY No $\rm T_{eff}$ & Torres No $\rm T_{eff}$ & SED No $\rm T_{eff}$ & $\sigma$ & Units	 \\
	\toprule
	$M_*$& $ 1.46^{+0.2}_{-0.16} $& $ 1.46^{+0.2}_{-0.13} $& $ 1.61^{+0.41}_{-0.24} $& $ 1.18^{+0.19}_{-0.16} $& $1.32$& $(\msun)$	\\
	$R_*$& $ 1.557^{+0.089}_{-0.064} $& $ 1.558^{+0.089}_{-0.054} $& $ 1.624^{+0.13}_{-0.1} $& $ 1.452^{+0.07}_{-0.067} $& $1.5$& $(\rsun)$	\\
	$L_*$& $ 4.0^{+2.2}_{-1.0} $& $ 4.3^{+2.4}_{-1.0} $& $ 6.2^{+8.5}_{-2.6} $& $ 3.58^{+0.62}_{-0.49} $& $0.47$& $(\lsun)$	\\
	$\rho_*$& $ 0.546^{+0.019}_{-0.028} $& $ 0.545^{+0.02}_{-0.028} $& $ 0.538^{+0.025}_{-0.029} $& $ 0.547^{+0.018}_{-0.027} $& $0.33$& (cgs)	\\
	$\log{g}$& $ 4.217^{+0.019}_{-0.019} $& $ 4.218^{+0.018}_{-0.017} $& $ 4.229^{+0.03}_{-0.023} $& $ 4.187^{+0.026}_{-0.029} $& $1.53$& (cgs)	\\
	$T_{\rm eff}$& $ 6550.0^{+570.0}_{-340.0} $& $ 6640.0^{+620.0}_{-350.0} $& $ 7170.0^{+1400.0}_{-710.0} $& $ 6590.0^{+320.0}_{-270.0} $& $0.59$& (K)	\\
	$[{\rm Fe/H}]$& $ 0.23^{+0.079}_{-0.079} $& $ 0.232^{+0.079}_{-0.079} $& $ 0.238^{+0.08}_{-0.08} $& $ 0.227^{+0.078}_{-0.078} $& $0.14$& (dex)	\\
\hline
	\end{tabular}

{\raggedright  $^{1}$ Optimal time of conjunction minimizes the covariance between $T_C$ and Period \par}
\end{table*}

\begin{table*}
	\centering
	\caption{Continuation of reesults for single constraint models without a spectroscopic \teff\ prior. The maximum difference between the models divided by the largest uncertainty is described by $\sigma$.}
	\label{tab:noTeffCanonicalCont}
	\begin{tabular}{lcccccl}
	\toprule	& MIST No $\rm T_{eff}$ prior & YY No $\rm T_{eff}$ & Torres No $\rm T_{eff}$ & SED No $\rm T_{eff}$ & $\sigma$ & Units	 \\
	\toprule
	$\rho_P$& $ 0.397^{+0.096}_{-0.096} $& $ 0.395^{+0.096}_{-0.097} $& $ 0.375^{+0.094}_{-0.091} $& $ 0.431^{+0.099}_{-0.1} $& $0.56$& Density (cgs)	\\
	$logg_P$& $ 3.065^{+0.088}_{-0.12} $& $ 3.063^{+0.088}_{-0.12} $& $ 3.059^{+0.088}_{-0.12} $& $ 3.066^{+0.088}_{-0.12} $& $0.07$& Surface gravity	\\
	$\Theta$& $ 0.046^{+0.011}_{-0.011} $& $ 0.046^{+0.011}_{-0.011} $& $ 0.044^{+0.01}_{-0.01} $& $ 0.05^{+0.011}_{-0.012} $& $0.52$& Safronov Number	\\
	$\fave$& $ 2.24^{+0.94}_{-0.44} $& $ 2.36^{+1.1}_{-0.47} $& $ 3.3^{+3.4}_{-1.1} $& $ 2.29^{+0.49}_{-0.37} $& $0.47$& \specialcell{Incident Flux \\ (\fluxcgs)}	\\
	$T_P$& $ 2458534.0884^{+0.00014}_{-0.00014} $& $ 2458534.0884^{+0.00014}_{-0.00014} $& $ 2458534.08839^{+0.00014}_{-0.00014} $& $ 2458534.08842^{+0.00014}_{-0.00014} $& $0.21$& \specialcell{Time of Periastron \\ (\bjdtdb)}	\\
	$T_S$& $ 2458535.75313^{+0.00014}_{-0.00014} $& $ 2458535.75313^{+0.00014}_{-0.00014} $& $ 2458535.75312^{+0.00014}_{-0.00014} $& $ 2458535.75316^{+0.00014}_{-0.00014} $& $0.18$& Time of eclipse (\bjdtdb)	\\
	$T_A$& $ 2458536.5855^{+0.00013}_{-0.00014} $& $ 2458536.5855^{+0.00014}_{-0.00014} $& $ 2458536.58549^{+0.00014}_{-0.00014} $& $ 2458536.58552^{+0.00014}_{-0.00014} $& $0.21$& \specialcell{Time of Ascending \\ Node (\bjdtdb)}	\\
\hline
	\end{tabular}
{\raggedright  $^{2}$ the velocity of the planet at the time of transit with the estimated eccentricity, Ve, divided by the velocity the planet would have if its orbit were circular, Vc \par}
\end{table*}

%%%%%%%%%%%%%%%%%%%%%%%%%%%%%%%%%%%%%%%%%%%%%%%%%%%%%%%%%%%%%%%%%%%%%%%%%%%%%%%%%%%%%%%%%%%%%%
\begin{table*}
	\centering
	\caption{Results for single constraint models without Claret tables priors on the Limb Darkening Coefficients The maximum difference between the models divided by the largest uncertainty is described by $\sigma$.}
	\label{tab:noClaretCanonical}
\begin{tabular}{lcccccl}
	\toprule	& MIST No Claret Tables & YY No Claret Tables & Torres No Claret Tables & SED No Claret Tables & $\sigma$ & Units	 \\
	\toprule
	$M_*$& $ 1.408^{+0.088}_{-0.089} $& $ 1.39^{+0.059}_{-0.057} $& $ 1.381^{+0.076}_{-0.073} $& $ 1.16^{+0.22}_{-0.18} $& $1.24$& $(\msun)$	\\
	$R_*$& $ 1.549^{+0.045}_{-0.043} $& $ 1.543^{+0.038}_{-0.035} $& $ 1.541^{+0.041}_{-0.038} $& $ 1.454^{+0.083}_{-0.077} $& $1.19$& $(\rsun)$	\\
	$L_*$& $ 3.67^{+0.44}_{-0.41} $& $ 3.68^{+0.42}_{-0.38} $& $ 3.67^{+0.42}_{-0.38} $& $ 3.59^{+0.79}_{-0.62} $& $0.13$& $(\lsun)$	\\
	$\rho_*$& $ 0.534^{+0.028}_{-0.029} $& $ 0.534^{+0.028}_{-0.029} $& $ 0.532^{+0.029}_{-0.029} $& $ 0.533^{+0.028}_{-0.03} $& $0.07$& (cgs)	\\
	$\log{g}$& $ 4.206^{+0.016}_{-0.018} $& $ 4.205^{+0.015}_{-0.017} $& $ 4.203^{+0.016}_{-0.017} $& $ 4.178^{+0.032}_{-0.033} $& $0.86$& (cgs)	\\
	$T_{\rm eff}$& $ 6410.0^{+130.0}_{-130.0} $& $ 6430.0^{+130.0}_{-130.0} $& $ 6430.0^{+130.0}_{-130.0} $& $ 6590.0^{+430.0}_{-390.0} $& $0.44$& (K)	\\
	$[{\rm Fe/H}]$& $ 0.237^{+0.079}_{-0.08} $& $ 0.24^{+0.08}_{-0.079} $& $ 0.24^{+0.08}_{-0.08} $& $ 0.23^{+0.079}_{-0.079} $& $0.12$& (dex)	\\
\hline
	\end{tabular}

{\raggedright  $^{1}$ Optimal time of conjunction minimizes the covariance between $T_C$ and Period \par}
\end{table*}

\begin{table*}
	\centering
	\caption{Continuation of results for single constraint models without Claret tables priors on the Limb Darkening Coefficients The maximum difference between the models divided by the largest uncertainty is described by $\sigma$.}
	\label{tab:noClaretCanonicalCont}
	\begin{tabular}{lcccccl}
	\toprule	& MIST - No Claret Tables & YY - No Claret Tables & Torres - No Claret Tables & SED N- No Claret Tables & $\sigma$ & Units	 \\
	\toprule
	$\rho_P$& $ 0.391^{+0.092}_{-0.094} $& $ 0.394^{+0.094}_{-0.096} $& $ 0.393^{+0.093}_{-0.094} $& $ 0.417^{+0.1}_{-0.1} $& $0.26$& Density (cgs)	\\
	$logg_P$& $ 3.054^{+0.089}_{-0.12} $& $ 3.055^{+0.09}_{-0.12} $& $ 3.054^{+0.089}_{-0.12} $& $ 3.055^{+0.089}_{-0.12} $& $0.01$& Surface gravity	\\
	$\Theta$& $ 0.046^{+0.01}_{-0.011} $& $ 0.046^{+0.01}_{-0.011} $& $ 0.046^{+0.01}_{-0.011} $& $ 0.049^{+0.011}_{-0.012} $& $0.26$& Safronov Number	\\
	$\fave$& $ 2.08^{+0.19}_{-0.18} $& $ 2.11^{+0.19}_{-0.18} $& $ 2.11^{+0.19}_{-0.18} $& $ 2.32^{+0.67}_{-0.51} $& $0.41$& \specialcell{Incident Flux \\ (\fluxcgs)}	\\
	$T_P$& $ 2458534.0884^{+0.00014}_{-0.00014} $& $ 2458534.0884^{+0.00014}_{-0.00014} $& $ 2458534.0884^{+0.00014}_{-0.00014} $& $ 2458534.08842^{+0.00014}_{-0.00014} $& $0.14$& \specialcell{Time of Periastron \\ (\bjdtdb)}	\\
	$T_S$& $ 2458535.75313^{+0.00014}_{-0.00014} $& $ 2458535.75313^{+0.00014}_{-0.00014} $& $ 2458535.75313^{+0.00014}_{-0.00014} $& $ 2458535.75315^{+0.00014}_{-0.00014} $& $0.11$& Time of eclipse (\bjdtdb)	\\
	$T_A$& $ 2458536.5855^{+0.00014}_{-0.00014} $& $ 2458536.5855^{+0.00014}_{-0.00014} $& $ 2458536.5855^{+0.00014}_{-0.00014} $& $ 2458536.58551^{+0.00014}_{-0.00014} $& $0.07$& \specialcell{Time of Ascending \\ Node (\bjdtdb)}	\\
\hline
	\end{tabular}
{\raggedright  $^{2}$ the velocity of the planet at the time of transit with the estimated eccentricity, Ve, divided by the velocity the planet would have if its orbit were circular, Vc \par}
\end{table*}

%%%%%%%%%%%%%%%%%%%%%%%%%%%%%%%%%%%%%%%%%%%%%%%%%%%%%%%%%%%%%%%%%%%%%%%%%%%%%%%%%%%%%%%%%%%%%%
\begin{table*}
	\centering
	\caption{Results for circular double constraint models. The maximum difference between the models divided by the largest uncertainty is described by $\sigma$.}
	\label{tab:CanoncialCombo}
	\begin{tabular}{lccccl}
	\toprule	& MIST +SED & YY + SED & Torres + SED & $\sigma$ & Units	 \\
	\toprule
	$M_*$& $ 1.366^{+0.072}_{-0.076} $& $ 1.364^{+0.05}_{-0.048} $& $ 1.382^{+0.074}_{-0.071} $& $0.24$& $(\msun)$	\\
	$R_*$& $ 1.516^{+0.031}_{-0.031} $& $ 1.515^{+0.026}_{-0.022} $& $ 1.528^{+0.037}_{-0.033} $& $0.37$& $(\rsun)$	\\
	$L_*$& $ 3.46^{+0.32}_{-0.29} $& $ 3.49^{+0.31}_{-0.29} $& $ 3.64^{+0.39}_{-0.36} $& $0.48$& $(\lsun)$	\\
	$\rho_*$& $ 0.556^{+0.011}_{-0.022} $& $ 0.556^{+0.012}_{-0.023} $& $ 0.549^{+0.017}_{-0.027} $& $0.32$& (cgs)	\\
	$\log{g}$& $ 4.213^{+0.011}_{-0.014} $& $ 4.2126^{+0.0085}_{-0.012} $& $ 4.21^{+0.012}_{-0.015} $& $0.22$& (cgs)	\\
	$T_{\rm eff}$& $ 6400.0^{+110.0}_{-110.0} $& $ 6410.0^{+110.0}_{-110.0} $& $ 6450.0^{+120.0}_{-120.0} $& $0.42$& (K)	\\
	$[{\rm Fe/H}]$& $ 0.213^{+0.076}_{-0.076} $& $ 0.213^{+0.079}_{-0.077} $& $ 0.236^{+0.079}_{-0.079} $& $0.29$& (dex)	\\
\hline
	\end{tabular}

{\raggedright  $^{1}$ Optimal time of conjunction minimizes the covariance between $T_C$ and Period \par}
\end{table*}

\begin{table*}
	\centering
	\caption{Continuation Results for circular double constraint models The maximum difference between the models divided by the largest uncertainty is described by $\sigma$.}
	\label{tab:CanoncialComboCont}
	\begin{tabular}{lccccl}
	\toprule	& MIST +SED & YY + SED & Torres + SED & $\sigma$ & Units	 \\
	\toprule
	$\rho_P$& $ 0.42^{+0.095}_{-0.1} $& $ 0.418^{+0.094}_{-0.099} $& $ 0.408^{+0.093}_{-0.096} $& $0.12$& Density (cgs)	\\
	$logg_P$& $ 3.072^{+0.087}_{-0.12} $& $ 3.071^{+0.087}_{-0.12} $& $ 3.065^{+0.087}_{-0.12} $& $0.07$& Surface gravity	\\
	$\Theta$& $ 0.048^{+0.011}_{-0.011} $& $ 0.047^{+0.01}_{-0.011} $& $ 0.047^{+0.01}_{-0.011} $& $0.09$& Safronov Number	\\
	$\fave$& $ 2.01^{+0.14}_{-0.13} $& $ 2.03^{+0.14}_{-0.13} $& $ 2.1^{+0.18}_{-0.16} $& $0.53$& \specialcell{Incident Flux \\ (\fluxcgs)}	\\
	$T_P$& $ 2458534.08841^{+0.00014}_{-0.00014} $& $ 2458534.08841^{+0.00014}_{-0.00014} $& $ 2458534.08841^{+0.00013}_{-0.00014} $& $0$& \specialcell{Time of Periastron \\ (\bjdtdb)}	\\
	$T_S$& $ 2458535.75314^{+0.00014}_{-0.00014} $& $ 2458535.75314^{+0.00014}_{-0.00014} $& $ 2458535.75314^{+0.00013}_{-0.00014} $& $0$& Time of eclipse (\bjdtdb)	\\
	$T_A$& $ 2458536.58551^{+0.00013}_{-0.00014} $& $ 2458536.58551^{+0.00014}_{-0.00014} $& $ 2458536.58551^{+0.00013}_{-0.00014} $& $0$& \specialcell{Time of Ascending \\ Node (\bjdtdb)}	\\
\hline
	\end{tabular}

{\raggedright  $^{2}$ the velocity of the planet at the time of transit with the estimated eccentricity, Ve, divided by the velocity the planet would have if its orbit were circular, Vc \par}
\end{table*}

%%%%%%%%%%%%%%%%%%%%%%%%%%%%%%%%%%%%%%%%%%%%%%%%%%%%%%%%%%%%%%%%%%%%%%%%%%%%%%%%%%%%%%%%%%%%%%

\begin{table*}
	\centering
	\caption{Results for eccentric double constraint models. The maximum difference between the models divided by the largest uncertainty is described by $\sigma$.}
	\label{tab:eccCanoncialCombo}
	\begin{tabular}{lccccl}
	\toprule	& MIST +SED & YY + SED & Torres + SED & $\sigma$ & Units	 \\	& Eccentric & Eccentric & Eccentric &  & 	 \\
	\toprule
	$M_*$& $ 1.372^{+0.071}_{-0.077} $& $ 1.369^{+0.05}_{-0.048} $& $ 1.355^{+0.065}_{-0.062} $& $0.23$& $(\msun)$	\\
	$R_*$& $ 1.492^{+0.054}_{-0.056} $& $ 1.488^{+0.052}_{-0.055} $& $ 1.493^{+0.053}_{-0.055} $& $0.09$& $(\rsun)$	\\
	$L_*$& $ 3.39^{+0.34}_{-0.31} $& $ 3.41^{+0.34}_{-0.3} $& $ 3.42^{+0.34}_{-0.3} $& $0.09$& $(\lsun)$	\\
	$\rho_*$& $ 0.579^{+0.071}_{-0.055} $& $ 0.584^{+0.072}_{-0.054} $& $ 0.572^{+0.072}_{-0.054} $& $0.19$& (cgs)	\\
	$\log{g}$& $ 4.227^{+0.035}_{-0.032} $& $ 4.228^{+0.035}_{-0.029} $& $ 4.221^{+0.037}_{-0.031} $& $0.21$& (cgs)	\\
	$T_{\rm eff}$& $ 6420.0^{+110.0}_{-110.0} $& $ 6440.0^{+110.0}_{-110.0} $& $ 6430.0^{+120.0}_{-110.0} $& $0.17$& (K)	\\
\hline
	\end{tabular}

{\raggedright  $^{1}$ Optimal time of conjunction minimizes the covariance between $T_C$ and Period \par}
\end{table*}

\begin{table*}
	\centering
	\caption{Continuation of results for for eccentric double constraint models. The maximum difference between the models divided by the largest uncertainty is described by $\sigma$.}
	\label{tab:eccCanoncialComboCont}
	\begin{tabular}{lccccl}
	\toprule	& MIST +SED & YY + SED & Torres + SED & $\sigma$ & Units	 \\	& Eccentric & Eccentric & Eccentric &  & 	 \\
	\toprule
	$\rho_P$& $ 0.45^{+0.11}_{-0.11} $& $ 0.45^{+0.11}_{-0.11} $& $ 0.44^{+0.11}_{-0.11} $& $0.09$& Density (cgs)	\\
	$logg_P$& $ 3.093^{+0.092}_{-0.12} $& $ 3.095^{+0.092}_{-0.12} $& $ 3.089^{+0.092}_{-0.12} $& $0.06$& Surface gravity	\\
	$\Theta$& $ 0.049^{+0.011}_{-0.012} $& $ 0.049^{+0.011}_{-0.012} $& $ 0.049^{+0.011}_{-0.012} $& $0$& Safronov Number	\\
	$\fave$& $ 1.93^{+0.17}_{-0.16} $& $ 1.95^{+0.17}_{-0.16} $& $ 1.97^{+0.18}_{-0.17} $& $0.23$& \specialcell{Incident Flux \\ (\fluxcgs)}	\\
	$T_P$& $ 2458531.59^{+0.49}_{-0.27} $& $ 2458531.61^{+0.51}_{-0.27} $& $ 2458531.57^{+0.48}_{-0.28} $& $0.1$& \specialcell{Time of Periastron \\ (\bjdtdb)}	\\
	$T_S$& $ 2458535.59^{+0.16}_{-0.22} $& $ 2458535.6^{+0.16}_{-0.21} $& $ 2458535.59^{+0.16}_{-0.21} $& $0.05$& Time of eclipse (\bjdtdb)	\\
\hline
	\end{tabular}

{\raggedright  $^{2}$ the velocity of the planet at the time of transit with the estimated eccentricity, Ve, divided by the velocity the planet would have if its orbit were circular, Vc \par}
\end{table*}

%%%%%%%%%%%%%%%%%%%%%%%%%%%%%%%%%%%%%%%%%%%%%%%%%%%%%%%%%%%%%%%%%%%%%%%%%%%%%%%%%%%%%%%%%%%%%%

 \begin{table*}
	\centering
	\caption{Results for double constraint models without a spectroscopic \teff\ prior. The maximum difference between the models divided by the largest uncertainty is described by $\sigma$.}
	\label{tab:noTeffCanoncialCombo}
	\begin{tabular}{lccccl}
	\toprule	& MIST +SED & YY + SED & Torres + SED & $\sigma$ & Units	 \\	& No \teff\ prior & No \teff\ prior & No \teff\ prior &  & 	 \\
	\toprule
	$M_*$& $ 1.353^{+0.082}_{-0.081} $& $ 1.356^{+0.068}_{-0.06} $& $ 1.345^{+0.073}_{-0.067} $& $0.13$& $(\msun)$	\\
	$R_*$& $ 1.512^{+0.033}_{-0.032} $& $ 1.513^{+0.029}_{-0.026} $& $ 1.51^{+0.031}_{-0.028} $& $0.09$& $(\rsun)$	\\
	$L_*$& $ 3.33^{+0.5}_{-0.4} $& $ 3.42^{+0.52}_{-0.43} $& $ 3.44^{+0.54}_{-0.43} $& $0.23$& $(\lsun)$	\\
	$\rho_*$& $ 0.556^{+0.012}_{-0.023} $& $ 0.556^{+0.012}_{-0.022} $& $ 0.554^{+0.013}_{-0.024} $& $0.11$& (cgs)	\\
	$\log{g}$& $ 4.211^{+0.012}_{-0.015} $& $ 4.212^{+0.01}_{-0.013} $& $ 4.21^{+0.011}_{-0.014} $& $0.15$& (cgs)	\\
	$T_{\rm eff}$& $ 6340.0^{+180.0}_{-160.0} $& $ 6380.0^{+180.0}_{-170.0} $& $ 6400.0^{+200.0}_{-180.0} $& $0.32$& (K)	\\
\hline
	\end{tabular}

{\raggedright  $^{1}$ Optimal time of conjunction minimizes the covariance between $T_C$ and Period \par}
\end{table*}

\begin{table*}
	\centering
	\caption{Continuation of results for double constraint models without a spectroscopic \teff\ prior. The maximum difference between the models divided by the largest uncertainty is described by $\sigma$.}
	\label{tab:noTeffCanoncialComboCont}
	\begin{tabular}{lccccl}
	\toprule	& MIST +SED & YY + SED & Torres + SED & $\sigma$ & Units	 \\	& No \teff\ prior & No \teff\ prior & No \teff\ prior &  & 	 \\
	\toprule
	$\rho_P$& $ 0.419^{+0.095}_{-0.099} $& $ 0.418^{+0.095}_{-0.099} $& $ 0.419^{+0.094}_{-0.098} $& $0.01$& Density (cgs)	\\
	$logg_P$& $ 3.071^{+0.088}_{-0.12} $& $ 3.07^{+0.088}_{-0.12} $& $ 3.07^{+0.087}_{-0.11} $& $0.01$& Surface gravity	\\
	$\Theta$& $ 0.048^{+0.011}_{-0.011} $& $ 0.047^{+0.011}_{-0.011} $& $ 0.048^{+0.011}_{-0.011} $& $0.09$& Safronov Number	\\
	$\fave$& $ 1.95^{+0.23}_{-0.19} $& $ 1.99^{+0.24}_{-0.2} $& $ 2.02^{+0.26}_{-0.21} $& $0.3$& \specialcell{Incident Flux \\ (\fluxcgs)}	\\
	$T_P$& $ 2458534.08841^{+0.00014}_{-0.00014} $& $ 2458534.08841^{+0.00013}_{-0.00014} $& $ 2458534.08841^{+0.00014}_{-0.00014} $& $0$& \specialcell{Time of Periastron \\ (\bjdtdb)}	\\
	$T_S$& $ 2458532.42367^{+0.00014}_{-0.00014} $& $ 2458532.42367^{+0.00014}_{-0.00014} $& $ 2458532.42368^{+0.00014}_{-0.00014} $& $0$& Time of eclipse (\bjdtdb)	\\
\hline
	\end{tabular}

{\raggedright  $^{2}$ the velocity of the planet at the time of transit with the estimated eccentricity, Ve, divided by the velocity the planet would have if its orbit were circular, Vc \par}
\end{table*}

%%%%%%%%%%%%%%%%%%%%%%%%%%%%%%%%%%%%%%%%%%%%%%%%%%%%%%%%%%%%%%%%%%%%%%%%%%%%%%%%%%%%%%%%%%%%%%

\begin{table*}
	\centering
	\caption{Results for double constraint models without Claret tables priors on the Limb Darkening Coefficients. The maximum difference between the models divided by the largest uncertainty is described by $\sigma$.}
	\label{tab:noClaretCanoncialCombo}
	\begin{tabular}{lccccl}
	\toprule	& MIST +SED & YY + SED & Torres + SED & $\sigma$ & Units	 \\	& No Claret Tables & No Claret Tables & No Claret Tables &  & 	 \\
	\toprule
	$M_*$& $ 1.357^{+0.074}_{-0.077} $& $ 1.361^{+0.05}_{-0.049} $& $ 1.328^{+0.075}_{-0.066} $& $0.44$& $(\msun)$	\\
	$R_*$& $ 1.521^{+0.035}_{-0.033} $& $ 1.522^{+0.03}_{-0.026} $& $ 1.517^{+0.033}_{-0.031} $& $0.15$& $(\rsun)$	\\
	$L_*$& $ 3.43^{+0.34}_{-0.31} $& $ 3.48^{+0.33}_{-0.3} $& $ 3.27^{+0.6}_{-0.43} $& $0.41$& $(\lsun)$	\\
	$\rho_*$& $ 0.545^{+0.021}_{-0.027} $& $ 0.546^{+0.021}_{-0.027} $& $ 0.539^{+0.025}_{-0.029} $& $0.26$& (cgs)	\\
	$\log{g}$& $ 4.207^{+0.014}_{-0.018} $& $ 4.208^{+0.012}_{-0.015} $& $ 4.201^{+0.016}_{-0.019} $& $0.4$& (cgs)	\\
	$T_{\rm eff}$& $ 6370.0^{+110.0}_{-110.0} $& $ 6390.0^{+110.0}_{-110.0} $& $ 6300.0^{+240.0}_{-190.0} $& $0.42$& (K)	\\
	$[{\rm Fe/H}]$& $ 0.217^{+0.077}_{-0.078} $& $ 0.219^{+0.078}_{-0.078} $& $ 0.224^{+0.078}_{-0.079} $& $0.09$& (dex)	\\
	$Age$& $ 2.03^{+1.1}_{-0.8} $& $ 1.97^{+0.59}_{-0.55} $& N/A& $0.06$& Age (Gyr)	\\
	$Period$& $ 3.32946476^{+7.7e-07}_{-7.5e-07} $& $ 3.32946477^{+7.7e-07}_{-7.7e-07} $& $ 3.32946477^{+7.7e-07}_{-7.7e-07} $& $0.01$& (days)	\\
	$R_P$& $ 1.418^{+0.037}_{-0.033} $& $ 1.419^{+0.034}_{-0.027} $& $ 1.416^{+0.035}_{-0.032} $& $0.09$& $(\rj)$	\\
	$M_P$& $ 0.94^{+0.21}_{-0.22} $& $ 0.94^{+0.21}_{-0.22} $& $ 0.93^{+0.21}_{-0.22} $& $0.05$& $(\mj)$	\\
	$T_C$& $ 2458534.0884^{+0.00014}_{-0.00014} $& $ 2458534.0884^{+0.00014}_{-0.00014} $& $ 2458534.0884^{+0.00014}_{-0.00014} $& $0$& \specialcell{Time of conjunction (\bjdtdb)}	\\
	$T_T$& $ 2458534.0884^{+0.00014}_{-0.00014} $& $ 2458534.0884^{+0.00014}_{-0.00014} $& $ 2458534.0884^{+0.00014}_{-0.00014} $& $0$& \specialcell{Time of minimum projected \\  separation (\bjdtdb)}	\\
	$T_{0}$& $  2458930.29471^{+0.0001}_{-0.0001} $& $ 2458930.29471^{+0.0001}_{-0.0001} $& $ 2458930.29471^{+0.0001}_{-0.0001} $& $0$& \specialcell{Optimal conjunction \\ Time$^{1}$ (\bjdtdb)}	\\
	$a$& $ 0.04833^{+0.00086}_{-0.00094} $& $ 0.04837^{+0.00059}_{-0.00058} $& $ 0.04798^{+0.00089}_{-0.00081} $& $0.43$& Semi-major axis (AU)	\\
	$i$& $ 88.42^{+0.94}_{-0.72} $& $ 88.44^{+0.92}_{-0.72} $& $ 88.25^{+0.96}_{-0.71} $& $0.23$& Inclination (Degrees)	\\
	$T_{eq}$& $ 1723.0^{+32.0}_{-31.0} $& $ 1728.0^{+31.0}_{-31.0} $& $ 1709.0^{+61.0}_{-50.0} $& $0.34$& (K)	\\
	$\tau_{\rm circ}$& $ 0.111^{+0.027}_{-0.027} $& $ 0.111^{+0.027}_{-0.027} $& $ 0.109^{+0.028}_{-0.027} $& $0.07$& \specialcell{Tidal circularization \\ timescale (Gyr)}	\\
	$K$& $ 104.0^{+23.0}_{-24.0} $& $ 104.0^{+23.0}_{-24.0} $& $ 104.0^{+23.0}_{-25.0} $& $0$& (m/s)	\\
	$R_P/R_*$& $ 0.09584^{+0.00051}_{-0.00044} $& $ 0.09584^{+0.00051}_{-0.00044} $& $ 0.09592^{+0.00054}_{-0.00048} $& $0.16$& Radius of planet in $R_{*}$	\\
	$a/R_*$& $ 6.839^{+0.088}_{-0.12} $& $ 6.842^{+0.085}_{-0.11} $& $ 6.81^{+0.1}_{-0.12} $& $0.29$& Semi-major axis in $R_{*}$	\\
	$\delta$& $ 0.009186^{+9.8e-05}_{-8.4e-05} $& $ 0.009184^{+9.7e-05}_{-8.4e-05} $& $ 0.009201^{+0.0001}_{-9.2e-05} $& $0.18$& Transit depth (fraction)	\\
	$Depth$& $ 0.01048^{+0.00038}_{-0.00035} $& $ 0.01048^{+0.00038}_{-0.00035} $& $ 0.01051^{+0.00039}_{-0.00035} $& $0.08$& Flux decrement at mid transit	\\
	$\tau$& $ 0.01529^{+0.00066}_{-0.00049} $& $ 0.01527^{+0.00065}_{-0.00047} $& $ 0.01542^{+0.00071}_{-0.00057} $& $0.23$& \specialcell{Ingress/egress duration (days)}	\\
	$T_{14}$& $ 0.16807^{+0.00053}_{-0.0005} $& $ 0.16807^{+0.00052}_{-0.0005} $& $ 0.16815^{+0.00054}_{-0.00052} $& $0.15$& Total transit duration (days)	\\
	$T_{FWHM}$& $ 0.15271^{+0.00053}_{-0.00053} $& $ 0.15272^{+0.00053}_{-0.00053} $& $ 0.15267^{+0.00055}_{-0.00055} $& $0.09$& FWHM transit duration (days)	\\
	$b$& $ 0.188^{+0.081}_{-0.11} $& $ 0.186^{+0.081}_{-0.11} $& $ 0.208^{+0.079}_{-0.11} $& $0.23$& Transit Impact parameter	\\
	$\delta_{S,2.5\mu m}$& $ 495.0^{+23.0}_{-21.0} $& $ 498.0^{+22.0}_{-21.0} $& $ 490.0^{+35.0}_{-29.0} $& $0.25$& \specialcell{Blackbody eclipse depth \\ at 2.5$\mu$m (ppm)}	\\
	$\delta_{S,5.0\mu m}$& $ 1214.0^{+34.0}_{-28.0} $& $ 1216.0^{+34.0}_{-28.0} $& $ 1214.0^{+39.0}_{-36.0} $& $0.05$& \specialcell{Blackbody eclipse depth \\ at 5.0$\mu$m (ppm)}	\\
	$\delta_{S,7.5\mu m}$& $ 1575.0^{+37.0}_{-29.0} $& $ 1576.0^{+37.0}_{-29.0} $& $ 1578.0^{+39.0}_{-35.0} $& $0.08$& \specialcell{Blackbody eclipse depth \\ at 7.5$\mu$m (ppm)}	\\
	\hline
	\end{tabular}

{\raggedright  $^{1}$ Optimal time of conjunction minimizes the covariance between $T_C$ and Period \par}
\end{table*}

\begin{table*}
	\centering
	\caption{Continuation of results for double constraint models without Claret tables priors on the Limb Darkening Coefficients. The maximum difference between the models divided by the largest uncertainty is described by $\sigma$.}
	\label{tab:noClaretCanoncialComboCont}
	\begin{tabular}{lccccl}
	\toprule	& MIST +SED & YY + SED & Torres + SED & $\sigma$ & Units	 \\	& No Claret Tables & No Claret Tables & No Claret Tables &  & 	 \\
	\toprule
	$\rho_P$& $ 0.407^{+0.093}_{-0.096} $& $ 0.407^{+0.093}_{-0.097} $& $ 0.404^{+0.094}_{-0.097} $& $0.03$& Density (cgs)	\\
	$logg_P$& $ 3.062^{+0.088}_{-0.12} $& $ 3.063^{+0.087}_{-0.12} $& $ 3.059^{+0.089}_{-0.12} $& $0.04$& Surface gravity	\\
	$\Theta$& $ 0.047^{+0.01}_{-0.011} $& $ 0.047^{+0.01}_{-0.011} $& $ 0.047^{+0.01}_{-0.011} $& $0$& Safronov Number	\\
	$\fave$& $ 2.0^{+0.15}_{-0.14} $& $ 2.02^{+0.15}_{-0.14} $& $ 1.94^{+0.29}_{-0.22} $& $0.31$& \specialcell{Incident Flux \\ (\fluxcgs)}	\\
	$T_P$& $ 2458534.0884^{+0.00014}_{-0.00014} $& $ 2458534.0884^{+0.00014}_{-0.00014} $& $ 2458534.0884^{+0.00014}_{-0.00014} $& $0$& \specialcell{Time of Periastron \\ (\bjdtdb)}	\\
	$T_S$& $ 2458535.75313^{+0.00014}_{-0.00014} $& $ 2458535.75313^{+0.00014}_{-0.00014} $& $ 2458535.75313^{+0.00014}_{-0.00014} $& $0$& Time of eclipse (\bjdtdb)	\\
	$T_A$& $ 2458536.5855^{+0.00014}_{-0.00014} $& $ 2458536.5855^{+0.00014}_{-0.00014} $& $ 2458536.5855^{+0.00014}_{-0.00014} $& $0$& \specialcell{Time of Ascending \\ Node (\bjdtdb)}	\\
	$T_D$& $ 2458534.92077^{+0.00014}_{-0.00014} $& $ 2458534.92077^{+0.00014}_{-0.00014} $& $ 2458534.92077^{+0.00014}_{-0.00014} $& $0$& \specialcell{Time of Descending \\ Node (\bjdtdb)}	\\
	$V_{e}/V_{c}$& $ 1.0^{+0.0}_{-0.0} $& $ 1.0^{+0.0}_{-0.0} $& $ 1.0^{+0.0}_{-0.0} $& $0$& Unitless$^{2}$	\\
	$M_P\sin i$& $ 0.94^{+0.21}_{-0.22} $& $ 0.94^{+0.21}_{-0.22} $& $ 0.93^{+0.21}_{-0.22} $& $0.05$& Minimum mass (\mj)	\\
	$d/R_*$& $ 6.839^{+0.088}_{-0.12} $& $ 6.842^{+0.085}_{-0.11} $& $ 6.81^{+0.1}_{-0.12} $& $0.29$& Separation at mid transit	\\
	$P_T$& $ 0.1322^{+0.0022}_{-0.0016} $& $ 0.1322^{+0.0022}_{-0.0016} $& $ 0.1327^{+0.0024}_{-0.0019} $& $0.23$& \specialcell{A priori non-grazing transit prob}	\\
	$P_{T,G}$& $ 0.1602^{+0.0028}_{-0.0021} $& $ 0.1602^{+0.0028}_{-0.002} $& $ 0.1608^{+0.0031}_{-0.0025} $& $0.21$& A priori transit prob	\\
	$u_{1}$& $ 0.21^{+0.21}_{-0.15} $& $ 0.22^{+0.21}_{-0.15} $& $ 0.22^{+0.21}_{-0.15} $& $0.06$& 	\\
	$u2_{0}$& $ 0.29^{+0.26}_{-0.31} $& $ 0.29^{+0.26}_{-0.31} $& $ 0.28^{+0.27}_{-0.31} $& $0.03$& 	\\
	$u1_{1}$& $ 0.53^{+0.29}_{-0.3} $& $ 0.53^{+0.29}_{-0.3} $& $ 0.54^{+0.29}_{-0.3} $& $0.03$& 	\\
	$u2_{1}$& $ -0.07^{+0.35}_{-0.24} $& $ -0.07^{+0.35}_{-0.24} $& $ -0.07^{+0.35}_{-0.24} $& $0$& 	\\
	$u1_{2}$& $ 0.257^{+0.057}_{-0.054} $& $ 0.257^{+0.057}_{-0.055} $& $ 0.26^{+0.058}_{-0.056} $& $0.05$& 	\\
	$u2_{2}$& $ 0.19^{+0.11}_{-0.11} $& $ 0.19^{+0.11}_{-0.11} $& $ 0.19^{+0.11}_{-0.12} $& $0$& 	\\
	$u1_{3}$& $ 0.45^{+0.28}_{-0.26} $& $ 0.45^{+0.28}_{-0.26} $& $ 0.44^{+0.29}_{-0.26} $& $0.04$& 	\\
	$u2_{3}$& $ 0.34^{+0.34}_{-0.42} $& $ 0.34^{+0.34}_{-0.42} $& $ 0.35^{+0.34}_{-0.42} $& $0.03$& 	\\
	${\gamma_{\rm rel}}_0$& $ 12195.0^{+20.0}_{-20.0} $& $ 12195.0^{+20.0}_{-20.0} $& $ 12195.0^{+20.0}_{-20.0} $& $0$& Relative RV Offset (m/s)	\\
	$\sigma_{J_0}$& $ 64.0^{+21.0}_{-14.0} $& $ 64.0^{+21.0}_{-14.0} $& $ 65.0^{+21.0}_{-14.0} $& $0.06$& RV Jitter (m/s)	\\
	${\sigma_J^2}_{0}$& $ 4200.0^{+3200.0}_{-1700.0} $& $ 4200.0^{+3100.0}_{-1600.0} $& $ 4200.0^{+3200.0}_{-1700.0} $& $0$& RV Jitter Variance	\\
	${\gamma_{\rm rel}}_{1}$& $ 12194.0^{+22.0}_{-21.0} $& $ 12194.0^{+22.0}_{-20.0} $& $ 12194.0^{+22.0}_{-20.0} $& $0$& 	\\
	${\sigma_J}_{1}$& $ 12.0^{+7.9}_{-12.0} $& $ 12.2^{+7.7}_{-12.0} $& $ 12.1^{+7.8}_{-12.0} $& $0.02$& 	\\
	${\sigma_J^2}_{1}$& $ 140.0^{+250.0}_{-330.0} $& $ 150.0^{+250.0}_{-330.0} $& $ 150.0^{+250.0}_{-330.0} $& $0.03$& 	\\
	$variance_0$& $ 1.47e-05^{+1.2e-06}_{-1.1e-06} $& $ 1.47e-05^{+1.2e-06}_{-1.1e-06} $& $ 1.47e-05^{+1.2e-06}_{-1.1e-06} $& $0$& 	\\
	$f0_0$& $ 1.00393^{+0.00022}_{-0.00022} $& $ 1.00393^{+0.00022}_{-0.00022} $& $ 1.00392^{+0.00022}_{-0.00022} $& $0.05$& Baseline flux	\\
	${C0_0}_0$& $ -0.01091^{+0.00069}_{-0.0007} $& $ -0.01091^{+0.00068}_{-0.00069} $& $ -0.0109^{+0.00068}_{-0.00069} $& $0.01$& Additive detrending coeff	\\
	${C1_0}_1$& $ 0.00104^{+0.00035}_{-0.00036} $& $ 0.00104^{+0.00035}_{-0.00035} $& $ 0.00104^{+0.00036}_{-0.00035} $& $0$& Additive detrending coeff	\\
	\hline
	\end{tabular}

{\raggedright  $^{2}$ the velocity of the planet at the time of transit with the estimated eccentricity, Ve, divided by the velocity the planet would have if its orbit were circular, Vc \par}
\end{table*}

\label{lastpage}
\end{document}